\documentclass[reprint,aps,prx,longbibliography]{revtex4-1}
\usepackage{color,graphicx,calc,url}
\usepackage{dcolumn}
\usepackage{bm}
\usepackage{amssymb}
\usepackage{amsmath}
\usepackage{wasysym}
\usepackage{mathrsfs} 
\usepackage{array}
\usepackage{mathtools}
\usepackage{commath}
\usepackage{adjustbox}
\usepackage{braket}
\usepackage{float}
\usepackage{textcomp}
\usepackage[colorlinks=true,
            linkcolor=red,
            urlcolor=blue,
            citecolor=blue]{hyperref}
            
\begin{document}
\title{Quadratic magnetooptic Kerr effect spectroscopy of Fe epitaxial films on MgO(001) substrates}
\author{Robin Silber$^{1,2,3}$, Ond\v{r}ej Stejskal$^{1,2}$, Luk\'a\v s Beran$^{4}$, Petr Cejpek$^{4}$, Roman Anto\v{s}$^4$, Tristan Matalla-Wagner$^{3}$, Jannis Thien$^5$, Olga Kuschel$^5$, Joachim Wollschl\"{a}ger$^5$, Martin Veis$^{4}$, Timo~Kuschel$^{3}$, Jaroslav Hamrle$^{4}$\email{Electronic mail: silber.rob@gmail.com}}
\affiliation{$^1$ Nanotechnology Centre, V\v{S}B-Technical University of Ostrava, 17. listopadu 15, 70833 Ostrava, Czech Republic\\
$^2$ IT4Innovations, V\v{S}B-Technical University of Ostrava, 17. listopadu 15, 70833 Ostrava, Czech Republic\\
$^3$ Center for Spinelectronic Materials and Devices, Department of Physics, Bielefeld University, Universit\"atsstra\ss e 25, 33615 Bielefeld, Germany\\
$^4$ Faculty of Mathematics and Physics, Charles University, Ke Karlovu 5, 12116 Prague, Czech Republic\\
$^5$ Department of Physics and Center of Physics and Chemistry of New Materials, Osnabr\"{u}ck University, 49076 Osnabr\"{u}ck, Germany}

\date{\today}


\begin{abstract}
The magnetooptic Kerr effect (MOKE) is a well known and handy tool to characterize ferro-, ferri- and antiferromagnetic materials. Many of the MOKE techniques employ effects solely linear in magnetization $\bm{M}$. Nevertheless, a higher-order term being proportional to $\bm{M}^2$ and called quadratic MOKE (QMOKE) can additionally contribute to the experimental data. Here, we present detailed QMOKE spectroscopy measurements in the range of 0.8 -- 5.5\,eV based  on a modified 8-directional method applied on ferromagnetic bcc Fe thin films grown on MgO substrates. From the measured QMOKE spectra, two further complex spectra of the QMOKE parameters $G_s$ and $2G_{44}$ are yielded. The difference between those two parameters, known as $\Delta G$, denotes the strength of the QMOKE anisotropy. Those QMOKE parameters give rise to the QMOKE tensor $\bm{G}$, fully describing the perturbation of the permittivity tensor in the second order in $\bm{M}$ for cubic crystal structures. We further present experimental measurements of ellipsometry and linear MOKE spectra, wherefrom permittivity in the zeroth and the first order in $\bm{M}$ are obtained, respectively. Finally, all those spectra are described by ab-initio calculations. 
\end{abstract}

\maketitle

\section{Introduction}

Ferromagnetic (FM) materials have been extensively studied due to their essential usage in data storage industry. Recently,  attention has been attracted to antiferromagnetic (AFM) materials due to the new possibility to control AFM spin orientation with an electrical current (based on spin-orbit torque effects) \cite{Jungwirth2016, Wadley2016}. Hence, there is increasing demand for fast and easily  accessible methods for AFM magnetization state characterization. However, most of the methods that are used for FM research are not applicable to AFM materials due to their lack of net magnetization. Nevertheless, the magnetooptic Kerr effect (MOKE) \cite{J.Kerr1877} and  magnetooptic (MO) effects in general, which are very powerful tools used in the field of FM research, can also be employed in  AFM research \cite{Nemec2018}. Although MOKE linear-in-magnetization (LinMOKE), being polar MOKE (PMOKE), longitudinal MOKE (LMOKE), and transversal MOKE (TMOKE) are only applicable to canted AFM and AFM dynamics \cite{Baierl2016,Kimel2009,Kampfrath2011}, it is the quadratic-in-magnetization part of the MOKE (QMOKE) that is employable for fully compensated AFM \cite{Saidl2017a}. 

There are several magnetooptic effects quadratic in magnetization. The QMOKE denotes a MO effect originating from non-zero off-diagonal reflection coefficients ($r_{sp}$ or $r_{ps}$), which appear due to the off-diagonal permittivity tensor elements (such as $\varepsilon_{xy}$). On the other hand, magnetic linear dichroism (MLD) and birefringence (also called the Voigt or Cotton-Mouton effect) denote MO effects observed in materials where different propagation and absorption of two linearly polarized modes occur, one being parallel and the other perpendicular to the magnetization vector $\bm{M}$ (or antiferromagnetic (N\'eel) vector $\bm{L}$ in the case of AFM). These effects originate from different diagonal elements of the permittivity tensor (for example $\varepsilon_{xx}-\varepsilon_{yy}$ for light propagating along the $z$-direction in the isotropic sample with in-plane $\bm{M}$ when $M_x\neq M_y$).

A more comprehensive approach to QMOKE is available, which also takes into account the anisotropy of QMOKE effects. Individual contributions to QMOKE can be measured and analyzed, stemming from the quadratic MO tensor $\bm{G}$ \cite{Visnovsky1986}, which describes a change to the permittivity tensor of the crystal in the second order in $\bm{M}$. The separation algorithm (known as the 8-directional method) has been developed for cubic (001) oriented crystals \cite{Postava2002}. It is based on MOKE measurement under 8 different $\bm{M}$ directions for different sample orientations with respect to the plane of incidence. Although applying this method on AFMs would be considerably challenging (because magnetic moments of AFMs have to be reoriented to desired directions), it is not in principle impossible. Switch of AFM through inverse MO effects is possible \cite{Kimel2009, Kanda2011} and control of AFM domain distribution was demonstrated by polarization-dependent optical annealing \cite{Higuchi2016}. But the easiest way to apply above mentioned separation process on AFMs would be to use easy-plane AFMs such as NiO(111) where sufficiently large magnetic field will align the moments perpendicular to the field direction due to Zeeman energy reduction by a small canting of the moments \cite{Hoogeboom2017}. Furthermore, if we take advantage of exchange coupling to an adjacent FM layer, e.g. Y$_3$Fe$_5$O$_{12}$ (YIG) \cite{Lin2017, Hou2017}, the requirement on the field strength will be substantially lower.

Nevertheless, to employ QMOKE measurements on a regular basis, the underlying origin of QMOKE must be well understood. Although  QMOKE effects have been already studied, especially in the case of Heusler compounds \cite{Hamrle07a, Hamrle07b, Muduli08, Muduli09, Trudel10c, Trudel11, Gaier2008, Wolf2011}, all the studies employ single wavelengths only. The MOKE spectroscopy together with ab-initio calculations is an appropriate combination to gain a good understanding of the microscopic origin of MO effects. In the field of LinMOKE spectroscopy, much work has already been done, \cite{Ferguson69, Krinchik68, Oppeneer92, Visnovsky1995, Uba1996, Visnovsky1999, Buschow2001, Hamrle2001, Hamrle2002, Grondilova2002, Visnovsky2005, Veis2014}, but in the field of QMOKE spectroscopy, only few systematic studies have been done so far \cite{Sepulveda2003,Lobov2012} where non of those studies is based on our approach to QMOKE spectroscopy.

Our separation process of different QMOKE contributions is stemming from 8-directional method \cite{Postava2002}, but we use a combination of just 4 directions and a sample rotation by 45$^\circ$ as will be described later in the text. This approach allows us to isolate QMOKE spectra that stem mostly from individual MO parameters and thus subsequently determine spectral dependencies of $G_s=G_{11}-G_{12}$ and $2G_{44}$, denoting MO parameters quadratic in $\bm{M}$ \cite{Hamrle07b, Buchmeier2009, Kuschel2011, Hamrlova2013}. Therefore, we start our study on FM bcc Fe thin films grown on MgO(001) substrates to get a basic understanding of QMOKE spectroscopy for further studies of AFMs. In the case of FM materials, we can simply orient the direction of $\bm{M}$ by using a sufficiently large external magnetic field and then separate different QMOKE contributions. We present a careful and detailed study of the MO parameters yielding process, and discuss all the experimental details that have to be considered in the process. We also present a comparison to the ab-initio calculations and values that have been reported in the literature so far. Possible sources of deviations between reported values are discussed.

Note that speaking of MOKE in general within this paper, we understand effects in the extended visible spectral range. There is a vast number of other magnetotransport phenomena in different spectral ranges.  In the dc spectral range we can mention the well known anomalous Hall effect \cite{Nagaosa2010}, being linear in $\bm{M}$, and anisotropic magnetoresistance (AMR) \cite{AMR38} together with the planar Hall effect, both being quadratic in $\bm{M}$. Recently, in the terahertz region, an MO effect of free carriers, the so-called optical Hall effect \cite{Kuhne2014}, has also received much attention \cite{Chochol2016}. From the x-ray family there is the well known x-ray magnetic circular (linear) dichroism and birefringence, being linear (quadratic) in $\bm{M}$ \cite{Valencia2010, Mertins2001}. All those (and other) effects (together with LinMOKE and QMOKE) can be described by equal symmetry arguments, predicting the permittivity tensor contributions of the first and second order in $\bm{M}$ \cite{Visnovsky1986}. The same argumentation is valid for other transport phenomena induced, e.g., by heat. Here, thermomagnetic effects such as the anomalous Nernst effect (linear in $\bm{M}$)\cite{Ettings1886,Huang2011,Meier2013a} and the anisotropic magnetothermopower together with the planar Nernst effect (quadratic in $\bm{M}$)\cite{Schmid2013,Meier2013b,Reimer2017} define the thermopower (or Seebeck) tensor.

In the upcoming section~\ref{theory_moke}, a brief introduction to the theory of linear and quadratic MOKE is presented. In section~\ref{sample_character} we describe the sample preparation together with structural and magnetic characterization. Section~\ref{optical_character} provides the optical characterization and section~\ref{MO_char} the MO characterization of the samples, being LinMOKE and QMOKE spectroscopy together with QMOKE anisotropy measurements. Finally, in section~\ref{comparison_abini} we compare our experimental findings with ab-initio calculations and the literature. 

%%%%%%%%%%%%%%%%%%%%%%%%%%%%%%%%%%%%%%%%%%%
%-----THEORY OF LINEAR AND QUADRATIC MOKE
%%%%%%%%%%%%%%%%%%%%%%%%%%%%%%%%%%%%%%%%%%%
\section{Theory of linear and quadratic MOKE}
\label{theory_moke}

The complex Kerr angle $\Phi_{s/p}$ for $s$ and $p$ polarized incident light is defined as \cite{Visnovsky06, Hecht2002}

\begin{equation}
\begin{split}	
\Phi_{s}&{}=
-\frac{r_{ps}}{r_{ss}}=
\frac{\tan{\theta_{s}}+i\tan{\epsilon_{s}}}{1-i\tan{\theta_{s}}\tan{\epsilon_{s}}}\approx\theta_{s}+i\epsilon_{s}\quad,
\\[5mm]
\Phi_{p}&{}=\frac{r_{sp}}{r_{pp}}=
\frac{\tan{\theta_{p}}+i\tan{\epsilon_{p}}}{1-i\tan{\theta_{p}}\tan{\epsilon_{p}}}\approx\theta_{p}+i\epsilon_{p}
\quad.
\end{split}
\label{Kerr_basic}
\end{equation}

\noindent
Here, $\theta_{s/p}$ and $\epsilon_{s/p}$ are Kerr rotation and Kerr ellipticity, respectively. As $|{\Phi_{s/p}}|<1^{\circ}$ for transition metals \cite{Buschow2001}, one can use small angle approximation in Eq.~(\ref{Kerr_basic}).  The reflection coefficients $r_{ss}, r_{ps}, r_{sp}, r_{pp}$ are the elements of the reflection matrix $\bm{R}$ of the sample described by the Jones formalism \cite{Hecht2002} as

\begin{equation}
	\bm{R}=
	\begin{bmatrix}
		r_{ss} & r_{sp}\\
		r_{ps} & r_{pp}
	\end{bmatrix}.
\end{equation} 

\noindent
These reflection coefficients fundamentally depend on the permittivity tensor $\bm{\varepsilon}$ (second rank 3$\times$3 tensor) of the magnetized crystal \cite{Visnovsky06}. Elements $\varepsilon_{ij}$ of the permittivity tensor are complex-valued functions of photon energy, and its real and imaginary part corresponds to dispersion and absorption of the material, respectively. Changes in the permittivity tensor with  $\bm{M}$ can be described through the Taylor series: $\bm{\varepsilon}=\bm{\varepsilon}^{(0)}+\bm{\varepsilon}^{(1)}+\bm{\varepsilon}^{(2)}+... $ , where the superscript denotes the order in $\bm{M}$. In our work we ignore all the contributions of third and higher orders in $\bm{M}$, expressing the elements of the permittivity tensor $\bm{\varepsilon}$ as

\begin{equation}
	\varepsilon_{ij}=\varepsilon_{ij}^{(0)} \, + \, \underbrace{K_{ijk}M_k}_{{\varepsilon_{ij}^{(1)}\rightarrow\text{LinMOKE}}} \, + \,\, \underbrace{G_{ijkl} M_k M_l}_{\varepsilon_{ij}^{(2)}\rightarrow \text{QMOKE}}\,,
\label{rozepsana permitivita}
\end{equation}

\noindent
where $M_k$ and $M_l$ are the components of the normalized $\bm{M}$. $K_{ijk}$ and $G_{ijkl}$ are the components of the so-called linear and quadratic MO tensors $\bm{K}$ and $\bm{G}$ of the third and fourth rank, respectively \cite{Visnovsky1986}. In Eq.~(\ref{rozepsana permitivita}), the Einstein summation convention is used. Thus, the permittivity $\bm{\varepsilon}$ up to the second order in $\bm{M}$ is fully described. The general shape of $\bm{K}$ and $\bm{G}$ can be substantially simplified using the Onsager relation $\varepsilon_{ij}(\bm{M})=\varepsilon_{ji}(-\bm{M})$ and symmetry arguments of the material \cite{Visnovsky1986}. The form of these tensors for all crystallographic classes  was thoroughly studied by \v{S}tefan Vi\v{s}\v{n}ovsk\'y \cite{Visnovsky06}. A cubic crystal structure with inversion symmetry (e.g. bcc Fe as investigated in this work) simplifies the permittivity tensor by
 
\begin{subequations}
\begin{align}
 	\varepsilon_{ij}^{(0)} &{}= \,\delta_{ij}\varepsilon _{d},\\[1mm]
 	K_{ijk} &{}=\, \epsilon_{ijk}K, \\[1mm]
 	G_{iiii} &{}=\, G_{11}, \\[1mm]
 	G_{iijj} &{}=\, G_{12}, \qquad i\neq j,\\[1mm]
 	G_{1212} &{}=\, G_{1313}=G_{2323}=G_{44},
\end{align}
\end{subequations} 
 
\noindent
with  $\delta_{ij}$ and $\epsilon_{ijk}$ being the Kronecker delta and the Levi-Civita symbol, respectively. Hence, ${\varepsilon}^{(0)}_{ij}$ is a diagonal tensor described by a scalar $\varepsilon_d$ for each photon energy. The linear MO tensor $\bm{K}$ is described by one free parameter $K$ whereas the quadratic MO tensor $\bm{G}$ is determined by two free parameters $G_s=(G_{11}-G_{12})$ and $2G_{44}$. In the literature $\Delta G=G_s-2G_{44}$ is also used \cite{Hamrle07b, Kuschel2011}, denoting the anisotropic strength of the $\bm{G}$ tensor. The shape of these tensors for cubic crystals and its dependence on the crystal orientation are intensively discussed in the  literature \cite{Hamrlova2013, Visnovsky1986}. The physical meaning of $G_s$ and $2G_{44}$ is the following: $G_s$, $2G_{44}$ denote magnetic linear dichroism when magnetization is along the $\braket{100}$ and $\braket{110}$ directions, respectively. Namely, $G_s=\varepsilon_\parallel-\varepsilon_\perp$ for $\bm{M}\parallel \braket{100}$ and $2G_{44}=\varepsilon_\parallel-\varepsilon_\perp$ for $\bm{M}\parallel \braket{110}$, where the parallel ($\parallel$) and perpendicular ($\perp$) symbols denote the directions of linear light polarization (i.e.\ applied electric field) with respect to the magnetization direction, respectively \cite{Hamrlova2013}.

\begin{figure}
\begin{center}
\includegraphics{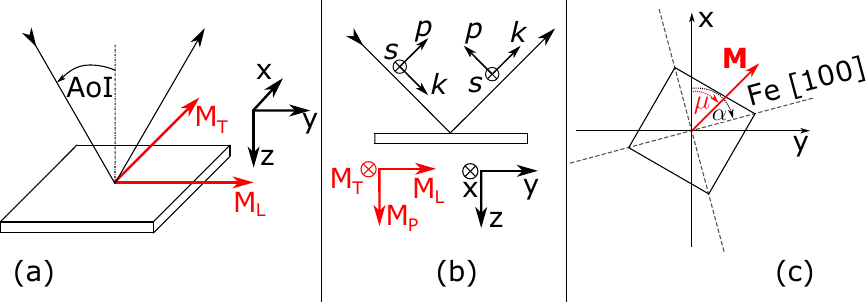}
\end{center}
\caption{(a) The right-handed coordinate system $\hat{x}$, $\hat{y}$, $\hat{z}$ is established  with respect to the plane of incidence and surface of the sample. Components of the in-plane normalized magnetization $M_{T}$ and $M_{L}$ are defined along the axes $\hat{x}$ and $\hat{y}$ of the coordinate system, respectively. (b) Definition of the right-handed Cartesian system $\hat{s}$, $\hat{p}$, $\hat{k}$ of incident and reflected beam. (c) Definition of positive in-plane rotation of the sample and magnetization within the $\hat{x}$, $\hat{y}$, $\hat{z}$ coordinate system, described by angle $\alpha$ and $\mu$, respectively.}
\label{xyz_sys}
\end{figure}

Let us briefly introduce the most important sign conventions. All definitions are based on a right-handed $\hat{x}$, $\hat{y}$, $\hat{z}$ coordinate system as sketched in Fig.~\ref{xyz_sys} with the $\hat{z}$-axis being normal to the sample surface (i.e.\ along Fe[001]) and pointing into the sample. The $\hat{y}$-axis is  parallel with the plane of light incidence and with the sample surface, while its positive direction is defined by the direction of $k_y$, being the $\hat{y}$-component of the wave vector of incident light. The orientation of the sample is then described by an angle $\alpha$, being the angle between the Fe [100] direction and the $\hat{x}$-axis of the coordinate system. Transverse, longitudinal and polar components of the normalized magnetization $M_{T}$, $M_{L}$ and $M_{P}$ are defined along the $\hat{x}$, $\hat{y}$ and $\hat{z}$ axes, respectively. Further, sign conventions are discussed in Appendix \ref{app_conv}. 

The analytical approximation for FM layers relating MOKE with the permittivity of the layer is \cite{Hamrle07b}

\begin{equation}
\begin{split}
\Phi_s &{}=-\frac{r_{ps}}{r_{ss}}= A_{s}\left(\varepsilon_{yx}-\frac{\varepsilon_{yz}\varepsilon_{zx}}{\varepsilon_d}\right)+B_s\varepsilon_{zx},\\[5mm]
\Phi_p &{}=\frac{r_{sp}}{r_{pp}}=-A_{p}\left(\varepsilon_{xy}-\frac{\varepsilon_{zy}\varepsilon_{xz}}{\varepsilon_d}\right)+B_p\varepsilon_{xz},
\end{split}
\label{Kerr_analyt}
\end{equation}

\noindent
with the weighting optical factors $A_{s/p}$ and $B_{s/p}$ being even and odd functions of the angle of incidence (AoI), respectively. 

In the following, we limit ourselves to in-plane normalized magnetization
\begin{equation}
 \frac{\bm{M}}{\norm{\bm{M}}}=
 \begin{bmatrix}
 M_T\\
 M_L\\
 0	
 \end{bmatrix}
 =
 \begin{bmatrix}
 	\cos\mu \\
 	\sin\mu \\
 	0
 \end{bmatrix},	
\label{Minplane}
\end{equation}

\noindent
where $\mu$ is the angle between the $\bm{M}$ direction and $\hat{x}$-axis of the coordinate system (see Fig.~\ref{xyz_sys}). From Eqs.~(\ref{rozepsana permitivita})--(\ref{Minplane}), the dependence of  $\Phi_{s/p}$  on $K$, $G_s$, $2G_{44}$ and on the angles $\alpha$ and $\mu$ can be derived as \cite{Hamrle07b, Kuschel2011, Silber2014} 

\begin{equation}
\begin{split}	
\Phi_{s/p} =&{}\pm A_{s/p}\left\{\frac{2G_{44}}{4}\left[\left(1+\cos{4\alpha}\right)\sin{2\mu}-\sin{4\alpha}\cos{2\mu}\right]\right.  \\[4mm]
&{}\qquad\quad+\frac{G_{s}}{4}\left.\left[\left(1-\cos{4\alpha}\right)\sin{2\mu}+\sin{4\alpha}\cos{2\mu}\right]\frac{}{}\right\} \\[4mm]
&{}\mp A_{s/p}\,\,\frac{K^2}{2\varepsilon_d}\sin{2\mu} \\[4mm]
&{}\pm B_{s/p}\,\,K\sin{\mu}.
\end{split}
\label{KerrDepend}
\end{equation}

\noindent
A change of the sign $\pm$ is related to the incident $s/p$ polarized light beam. From this expression, measurement sequences providing MOKE spectra originating mostly from individual MO parameters are developed \cite{Postava2002} and presented in Section \ref{MO_char}. 

%%%%%%%%%%%%%%%%%%%%%%%%%%%%%%%%%%%%%%%%%%%
%---PREPARATION AND STRUCTURAL CHARACTERIZATION
%%%%%%%%%%%%%%%%%%%%%%%%%%%%%%%%%%%%%%%%%%%
\section{Preparation, structural and magnetic anisotropy characterization of the samples}
\label{sample_character}

A series of epitaxial bcc Fe(001) thin films with various thicknesses were prepared in an Ar atmosphere of 2.1$\cdot 10^{-6}$\,bar using magnetron sputtering. The Fe layer was directly grown on the MgO(001) substrate with a growth rate of 0.25~nm/s. To prevent oxidation, the Fe layer was capped with approximately 2.5\,nm of silicon under the same conditions and with a growth rate of 0.18~nm/s. A reference sample of the MgO substrate with only silicon capping was prepared in order to determine the optical parameters of the capping layer independently. The sample set contains 10 samples with a nominal thicknesses of the Fe layer ranging from 0 nm to 30 nm as shown in Tab.~\ref{TabXRR}. Furthermore, an additional set of Fe samples grown by molecular beam epitaxy (MBE) on MgO(001) substrates and capped with Si were prepared to investigate the influence of the deposition process on the magnetooptic properties of Fe. Their preparation and comparison with the sputtered samples is discussed in Appendix~\ref{MBE_sample}.

\begin{figure}
\begin{center}
\includegraphics{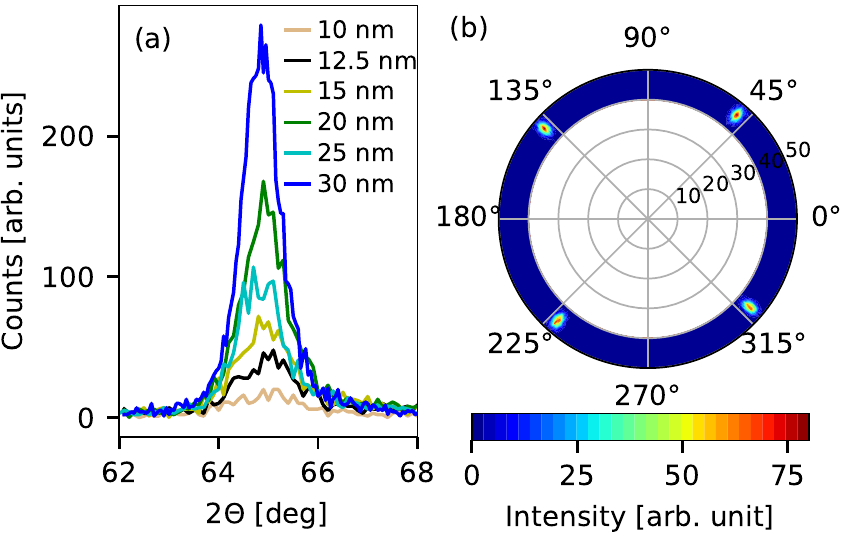}
\end{center}
\caption{(a) XRD $\Theta$ -- $2\Theta$ scans of the samples with a nominal thickness $>$ 10\,nm. Thinner samples do not provide sufficient peak intensity. (b) An off-specular XRD scan (Euler's cradle texture map) is presented for the Fe$\{$110$\}$ peaks at 2$\Theta$=44.738$^\circ$ of the sample with a nominal thickness of 20\,nm. The measurement was performed for full 360$^{\circ}$ sample rotation (angular axis of the plot) with the tilt of the sample $\Psi$=$\langle$40$^\circ$,50$^\circ$$\rangle$ (radial axis of the plot).}
\label{XRD}
\end{figure}
 
To verify crystallographic ordering and quality, Phillips X'pert Pro MPD PW3040-60 using a Cu-K$_{\alpha}$ source was employed. X-ray diffraction (XRD) $\Theta$ -- $2\Theta$ scans were performed around $2\Theta$ = 65$^\circ$, which is the position of the characteristic Fe(002) Bragg peak. Thinner samples provide very weak peaks due to the lack of the material in the thin layers as presented in Fig.~\ref{XRD}(a). Furthermore, for the sample with a nominal thickness of 20\,nm,  an off-specular texture mapping was performed using a Euler cradle (Fig.~\ref{XRD}(b)). During this scan the Fe$\{$110$\}$ peak at $2\Theta=44.738^\circ$ was used and we scanned $\Psi$ in the range of $40-50^\circ$ with full $360^\circ$ rotation of $\varphi$, where $\Psi$ and $\varphi$ are the tilt angle of the Euler cradle and the rotation angle of the sample around its surface normal, respectively. The result implies that the Fe layer within the sample is of good crystalline quality, showing a diffraction pattern in four-fold symmetry. 

\begin{figure}
\begin{center}
\includegraphics{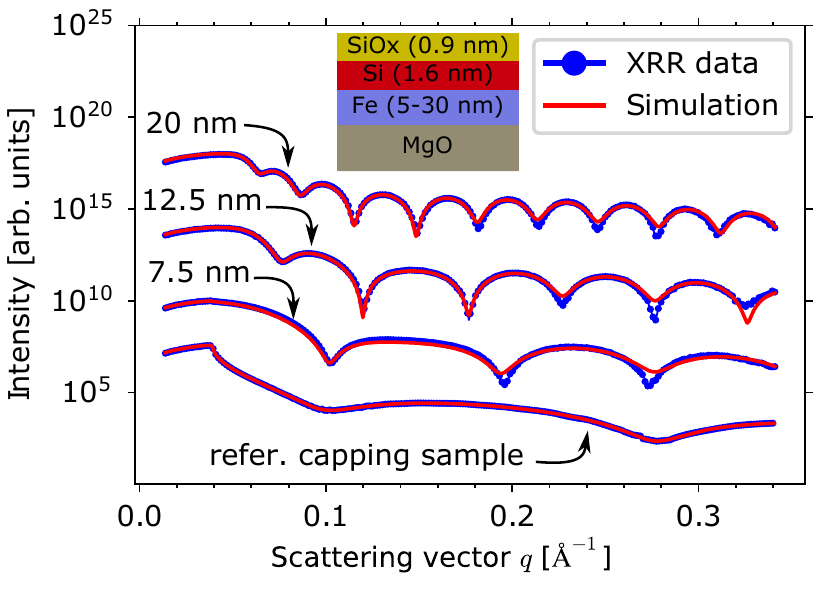}
\end{center}
\caption{Selected XRR scans (blue dots) and their simulation (red line) for several samples from the series. The periodicity of oscillations is well described, providing us with reliable information about the thickness of the layers in the samples. The damping of oscillations is low, suggesting low roughness of the interfaces within the samples. The curves are shifted vertically for clarity.}
\label{xrr_RSf}
\end{figure}

\begin{table}
	\begin{center}
		\begin{tabular}{crcrcrcrcrc}
			\hline
			\hline
			Nominal       && $d_{\mathrm{Fe}}$ &&$d_{\mathrm{cap}}$&& $\sigma_{\mathrm{MgO}}$&& $\sigma_{\mathrm{Fe}}$&& $\sigma_{\mathrm{cap}}$\\
			 thickness\,[nm]          && [nm] &&[nm]&&[nm]&& [nm]&& [nm]\\
			\hline
			 0.0\,nm			 && -  && 3.4  && 0.2 && - &&  0.3  \\
			 2.5\,nm            && 2.5 && 2.1  && 0.4 && 0.0 && 0.0  \\
			 5.0\,nm			 && 4.7 && 2.4  && 0.0 && 0.4 && 0.2  \\
			 7.5\,nm            && 6.9 && 2.5  && 0.0 && 0.3 &&  0.4  \\
			 10.0\,nm			 && 9.4 && 2.7  && 0.2 && 0.0 &&  0.6  \\
			 12.5\,nm            &&11.5 && 2.5  && 0.1 &&  0.3  &&  0.3  \\
			 15.0\,nm			 && 14.0 && 2.5  && 0.0 && 0.2 &&  0.1  \\
			 20.0\,nm            && 18.4 && 2.6  && 0.1 && 0.0  && 0.5  \\
			 25.0\,nm			 && 23.3 && 2.4  && 0.1 && 0.2 &&  0.6  \\
			 30.0\,nm            && 28.3 && 2.5 && 0.0 && 0.0  && 0.6  \\
		\hline
		\hline
		\end{tabular}
\caption{Thicknesses and roughnesses of the samples, as determined from XRR. Thicknesses of Fe layer $d_{\mathrm{Fe}}$ and capping layer $d_{\mathrm{cap}}$ are very robust parameters of the fit and the value for the error bars is $\pm$0.2\,nm. Although changes of the roughness $\sigma$ in tens of percent provide insignificant change to the result, it is clear that the roughness is very low as suggested in Fig.~\ref{xrr_RSf}, and the estimated value for the error bars is $\pm$0.5~nm.}
\label{TabXRR}
\end{center}
\end{table}

The thickness of each layer and the roughness of each interface was characterized by x-ray reflectivity (XRR) using the same diffractometer as for the XRD measurements. To analyze the XRR curves, the open-source program GenX \cite{Bjorck2007} based on the Parratt algorithm \cite{Parratt1954} was used. XRR scans are shown for selected samples in Fig.~\ref{xrr_RSf}. The periodicity of the oscillations is described very well by the model, providing reliable information about the thickness values of the Fe layers $d_{\mathrm{Fe}}$ and the capping layers $d_{\mathrm{cap}}$. The densities of the layers were fixed parameters of the fit and all values were taken from the literature \cite{Haynes2014, Custer1994}. The thickness of the native silicon oxide could not be clearly determined by the XRR technique as Si and SiO$_x$ have very similar densities. Hence, the thickness of the oxide was estimated (0.9~nm) with respect to the growth dynamics of the native silicon oxide \cite{Mende1983}. Tab.~\ref{TabXRR} summarizes all the values of the thickness and roughness provided by the XRR data fit. 

\begin{figure}
\begin{center}
\includegraphics{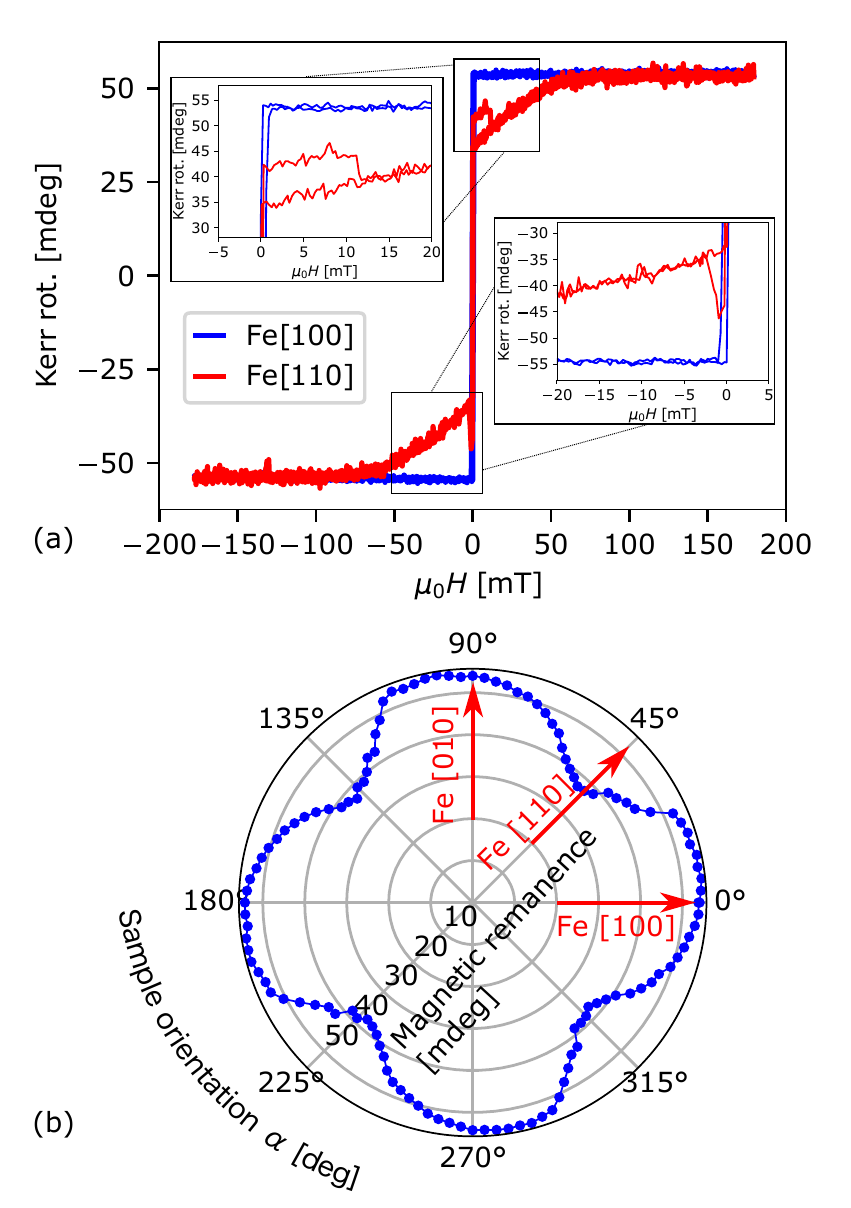}
\end{center}
\caption{Magnetic characterization of the sample with a nominal thickness of 12.5\,nm. (a) The LMOKE hysteresis loops at Fe[100] and Fe[110] external field directions. About 75\,mT is sufficient to saturate the sample in the in-plane hard axis. (b) In-plane magnetic remanence, with the in-plane magnetic easy and hard axes along Fe$\langle$100$\rangle$ and Fe$\langle$110$\rangle$ directions, respectively.}
\label{Loops_rem}
\end{figure} 

LMOKE hysteresis curves with an external magnetic field along Fe[100] and Fe[110] directions measured at $\lambda$=670\,nm (1.85\,eV) are shown in Fig.~\ref{Loops_rem}(a). The anisotropy of the magnetic remanence (an average value of positive and negative remanence) presented in Fig.~\ref{Loops_rem}(b) indicates the fourfold cubic magnetocrystalline anisotropy with the magnetic easy and hard axes along the Fe$\langle$100$\rangle$ and Fe$\langle$110$\rangle$ directions, respectively. Figure~\ref{Loops_rem}(b) further suggests that magnetic easy and hard axes are rotated slightly counter-clockwise with respect to Fe$\langle$100$\rangle$ and Fe$\langle$110$\rangle$ directions, respectively. This could be explained by a slight misalignment of the sample in the setup with respect to $\alpha=0^{\circ}$, but more probably by additional QMOKE contributions to the LMOKE loops as identified in the inset of Fig.~\ref{Loops_rem}(a). The magnetic field of $\approx$75\,mT is enough to saturate the sample in a magnetic in-plane hard axis, hence the in-plane magnetic field of 300\,mT used within QMOKE spectroscopy is more than sufficient to keep the sample saturated with any in-plane $\bm{M}$ direction. 

%%%%%%%%%%%%%%%%%%%%%%%%%%%%%%%%%%%%%%%%%%%
% OPTICAL CHARACTERIZATION
%%%%%%%%%%%%%%%%%%%%%%%%%%%%%%%%%%%%%%%%%%%
\section{Optical characterization}
\label{optical_character}

\begin{figure}
\begin{center}
\includegraphics{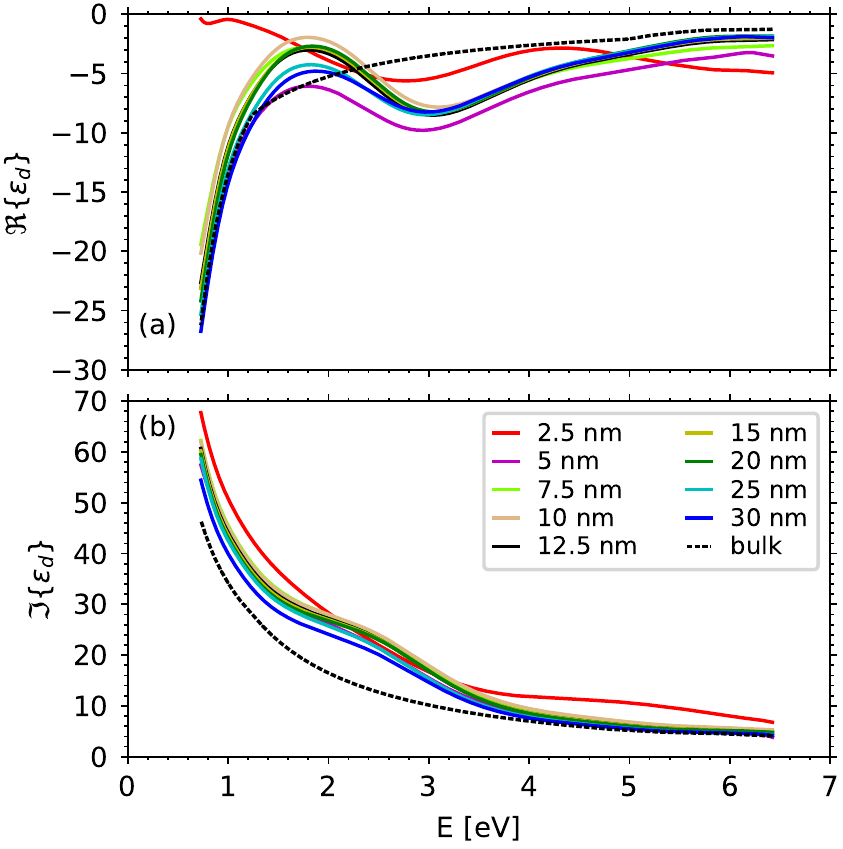}
\end{center}
\caption{The (a) real and (b) imaginary part of $\varepsilon_d$ of Fe layers. Black dashed lines are the $\varepsilon_d$ of Palik \cite{Palik} and were used as an initial guess for the fit of the $\varepsilon_d$ of the Fe layers for all the samples (full, coloured lines).}
\label{Elipso}
\end{figure}

The Mueller matrix ellipsometer Woolam RC2 was employed to determine spectral dependencies of $\varepsilon_d$ for all the layers within the investigated samples in the spectral range 0.7 -- 6.4\,eV. 
Spectra of $\varepsilon_d$ of Fe were determined by a multilayer optical model \cite{Yeh1980}, processed using CompleteEASE software \cite{J.A.WoollamCo.2008}. The thicknesses and roughnesses of the constituent layers were determined by XRR measurements. The permittivity of MgO and native SiO$_x$ was taken from the literature \cite{Palik}. From the measurement of the reference sample (MgO with the Si capping only, with nominal Fe thickness 0~nm), the permittivity of the Si layer was obtained. Hence, for all the remaining samples, $\varepsilon_d$ of the Fe layer was the only unknown and free variable of the fit.

The spectra of the imaginary part of $\varepsilon_d$ for Fe and Si layers were described by B-spline \cite{Johs2008}, while complementary spectra of the real part were determined through Kramers-Kronig relations. The B-spline is a fast and sturdy method for determining spectra of $\varepsilon_d$, but does not provide direct information about the electronic structure of the material. The resulting spectra of the real and imaginary part of Fe layers are presented in Figs.~\ref{Elipso} (a) and (b), respectively. The sample with a nominal thickness of 2.5~nm is deviating from the others, probably due to low crystallographic quality of the film. Although the characteristic peak at 2.5~eV in $\varepsilon_d$ imaginary spectra of the Fe layer is not present in the spectra of Fe by Palik \cite{Palik}, the position of this peak is consistent with other reports as shown in section~\ref{comparison_abini}.

%%%%%%%%%%%%%%%%%%%%%%%%%%%%%%%%%%%%%%%%%%%
%----MAGNETOOPTICAL CHARACTERIZATION
%%%%%%%%%%%%%%%%%%%%%%%%%%%%%%%%%%%%%%%%%%%
\section{Magnetooptic characterization}
\label{MO_char}

Three in-house built MOKE setups were employed to measure the LinMOKE and QMOKE response on the sample series. One setup (located at Bielefeld University) detects the MOKE with variation of the sample orientation $\alpha$ for a fixed photon energy 1.85~eV. Two other setups detect spectra of MOKE for a fixed sample orientation, measuring in the spectral range of 1.6 -- 4.9~eV (Charles University in Prague) and 1.2 -- 5.5~eV (Technical University of Ostrava), respectively, with perfect agreement of spectra obtained from both setups. The sample with a nominal thickness of 12.5~nm was later remeasured with an enhanced spectral range of 0.8 -- 5.5~eV. A detailed description of the spectroscopic setup at the University of Ostrava can be found in the literature \cite{Silber18}.

We now describe the QMOKE spectra measurement process. Using Eq.~(\ref{KerrDepend}) (describing the Kerr effect dependence on the angles $\alpha$, $\mu$ and the MO parameters $K$, $G_s$ and $2G_{44}$) we derive a measurement procedure separating MOKE contributions originating mostly from individual elements of the linear and quadratic MO tensors, $\bm{K}$ and $\bm{G}$, respectively. With the specified AoI and sample orientation $\alpha$, we measure MOKE with several in-plane $\bm{M}$ directions \cite{Postava2002}. To rotate $\bm{M}$ in the plane of the sample, a magnetic field of 300~mT is used and secures that the sample is always in magnetic saturation as proven in Fig.~\ref{Loops_rem}. Three MO contributions can be separated:

\begin{widetext}
\begin{subequations}
\begin{alignat}{4}
&\mathmakebox[\widthof{QQQQQQQQQQQQQQQQ}][l]{\mathrm{QMOKE}\sim G_s = Q_s:}&
\Phi_{s/p}^{\mu=45^{\circ}}+\Phi_{s/p}^{\mu=225^{\circ}}-\Phi_{s/p}^{\mu=135^{\circ}}-\Phi_{s/p}^{\mu=315^{\circ}}
&\mathmakebox[\widthof{EE}][c]{\approx}&
&\mathmakebox[\widthof{AAAAAAAAAAA}][l]{\pm\,2A_{s/p}\left(G_{s}-\frac{K^2}{\varepsilon_d}\right) ,}&
\begin{array}{r@{}l}
	&{}\alpha = 45^{\circ}\\
	&{}\mathrm{AoI}=5^{\circ}
\end{array}.
\label{QMOKE_sequence_s}
\\[3mm]
&\mathmakebox[\widthof{QQQQQQQQQQQQQQQQ}][l]{\mathrm{QMOKE}\sim 2G_{44} = Q_{44}:}&
\Phi_{s/p}^{\mu=45^{\circ}}+\Phi_{s/p}^{\mu=225^{\circ}}-\Phi_{s/p}^{\mu=135^{\circ}}-\Phi_{s/p}^{\mu=315^{\circ}}
&\mathmakebox[\widthof{EE}][c]{\approx}&
&\mathmakebox[\widthof{AAAAAAAAAAA}][l]{\pm\,2A_{s/p}\left(2G_{44}-\frac{K^2}{\varepsilon_d}\right) ,}& 
\begin{array}{r@{}l}
	&{}\alpha = 0^{\circ}\\
	&{}\mathrm{AoI}=5^{\circ}
\end{array}.
\label{QMOKE_sequence_44}
\\[3mm]
&\mathmakebox[\widthof{QQQQQQQQQQQQ}][l]{\mathrm{LMOKE}\sim K:}&
\Phi _{s/p}^{\mu=90^{\circ}}-\Phi _{s/p}^{\mu=270^{\circ}}
&\mathmakebox[\widthof{EE}][c]{\approx}&
&\mathmakebox[\widthof{AAAAAAAAAAA}][l]{\pm\,2B_{s/p}K ,}&
\begin{array}{r@{}l}
	&{}\alpha = \mathrm{arb.\,angle}\\ 
	&{}\mathrm{AoI}=45^{\circ}
\end{array}.
\label{LMOKE_sequence}
\end{alignat}
\end{subequations}
\end{widetext}

\noindent
where $\pm$ denotes $s/p$ MOKE effects. The AoI in the equations were chosen with respect to the AoI dependence of the optical weighting factors $A_{s/p}\sim\cos$(AoI) and $B_{s/p}\sim\sin$(AoI). Hence, the AoI in the Eqs.~(\ref{QMOKE_sequence_s})--(\ref{LMOKE_sequence}) only affects the amplitude of the acquired spectra and is not essential for the spectra separation process, unlike the sample orientation $\alpha$ and the magnetization directions $\mu$ that are vital to the measurement sequences. QMOKE and LMOKE spectra were measured at AoI=5$^\circ$ and 45$^\circ$, respectively. 

QMOKE and LMOKE measurement sequences are determined by Eqs.~(\ref{QMOKE_sequence_s}) -- (\ref{LMOKE_sequence}), left side, as a difference of MOKE effects for different magnetization orientations $\mu$ at specified sample orientation $\alpha$. We further use the denominations $Q_s$ and $Q_{44}$ for those QMOKE measurement sequences in Eqs.~(\ref{QMOKE_sequence_s}) and (\ref{QMOKE_sequence_44}), respectively. The right side of Eqs.~(\ref{QMOKE_sequence_s}) -- (\ref{LMOKE_sequence}) shows the outcome of those sequences when using the approximative description of MOKE, Eq.~(\ref{Kerr_analyt}), providing selectivity to $G_s$, $2G_{44}$ and $K$ within validity of Eq.~(\ref{Kerr_analyt}), respectively. 

The next step is to extract the MO parameters $G_s$, $2G_{44}$ and $K$ from the measured spectra using the phenomenological description of the MOKE spectra by Yeh's 4$\times$4 matrix formalism \cite{Visnovsky06, Yeh1980} based on classical Maxwell equations and boundary conditions. Propagation of coherent electromagnetic plane waves through a multilayer system is considered within this formalism. By solving the wave equation for each layer (characterized by its permittivity tensor and thickness), the reflection matrix $\bm{R}$ of the multilayer system can be obtained, which allows us to numerically calculate the MOKE angles of the sample according to Eq.~(\ref{Kerr_basic}). The thickness and the $\varepsilon_d$ of each layer is known from XRR and ellipsometry measurements, respectively. Nevertheless, the  permittivity tensor of the FM layer is described by the sum: $\bm{\varepsilon}=\bm{\varepsilon}^{(0)}+\bm{\varepsilon}^{(1)}+\bm{\varepsilon}^{(2)}$. Hence, $G_s$, $2G_{44}$ and $K$ are the unknowns in Yeh's 4$\times$4 matrix formalism calculations, being free parameters to the fit where both measured and calculated sequences are given by Eqs.~(\ref{QMOKE_sequence_s}) -- (\ref{LMOKE_sequence}), left side. However, as the measured and calculated spectra are determined by those equal sequences, the determination of spectra of the MO parameters $G_s$, $2G_{44}$ and $K$ is not affected by an approximation given by Eq.~(\ref{Kerr_analyt}). Finally, we would like to point out that the condition of proper positive direction of $\bm{M}$ rotation angle $\mu$ must be met. Although the opposite direction of $\bm{M}$ rotation will lead only to the opposite sign of experimental spectra, it may lead to completely incorrect spectra of $G_s$ and $2G_{44}$ parameters upon processing. We have checked that all sign conventions as defined in Appendix~\ref{app_conv} agree with experimental procedures, analytical descriptions, and numerical calculations. For further details about this issue, please see Appendix \ref{mag_dir}.

%%%%%%%%%%%%%%%%%%%%%%%%%%%%%%%%%%%%%%%%%%%
\subsection{QMOKE Anisotropy}
\label{QMOKE_anisotropy}

\begin{figure}
\begin{center}
\includegraphics{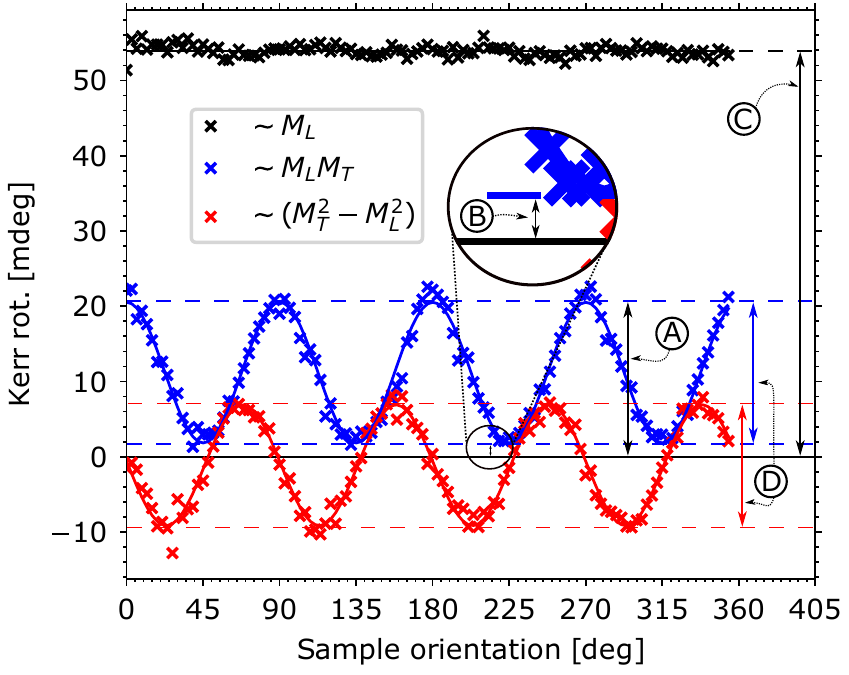}
\end{center}
\caption{QMOKE anisotropy measurement at a photon energy of 1.85\,eV, with AoI=45$^\circ$ for the sample with a nominal thickness of 12.5\,nm. The dependence of the three MOKE contributions, being LMOKE $\sim M_L$, QMOKE $\sim M_LM_T$ and QMOKE $\sim (M_T^2-M_L^2)$ on the sample orientation $\alpha$ is demonstrated. Further, several MOKE values are designated in the graph, being \raisebox{.0pt}{\textcircled{\raisebox{-.7pt} {A}}} = $A_{s}\left(2G_{44}-\frac{K^2}{\varepsilon_d}\right)$, \raisebox{.0pt}{\textcircled{\raisebox{-1.0pt} {B}}} = $A_{s}\left(G_{s}-\frac{K^2}{\varepsilon_d}\right)$, \raisebox{.0pt}{\textcircled{\raisebox{-1.0pt} {C}}} = $B_{s}K$ and \raisebox{.0pt}{\textcircled{\raisebox{-1.0pt} {D}}} = $A_{s}\Delta G$.}
\label{8_dir_method}
\end{figure}

The anisotropy of QMOKE is demonstrated by the so-called 8-directional method \cite{Postava2002}. The MOKE signal was detected for 8 in-plane magnetization directions, being $\mu=0^\circ+k\cdot 45^\circ, k=\{0,1,...,7\}$. From those measurements, constituent MOKE signals were separated, being namely LMOKE $\sim M_L$ contribution and two quadratic contributions QMOKE $\sim M_LM_T$, and QMOKE $\sim(M_L^2-M_T^2)$. Note that the separation process could be derived using Eq.~(\ref{KerrDepend}).

The dependences of those three MOKE contributions on the sample orientation $\alpha$ are yielded. In Fig.~\ref{8_dir_method} we present all three MOKE contributions measured for the sample with a nominal thickness of 12.5 nm at a photon energy of 1.85\,eV and with AoI=45$^\circ$. The fourfold anisotropy of the QMOKE contributions and isotropic LMOKE contribution follow the theory well (see $\alpha$ dependence in  Eq.~(\ref{KerrDepend})). Note that the separation processes of the contributions $\sim M_LM_T$ for $\alpha=45^\circ$,$0^\circ$ and $\sim M_L$ are identical as described in Eqs.~(\ref{QMOKE_sequence_s}) -- (\ref{LMOKE_sequence}), respectively.

%%%%%%%%%%%%%%%%%%%%%%%%%%%%%%%%%%%%%%%%%%%
\subsection{Linear MOKE spectroscopy}

\begin{figure}
\begin{center}
\includegraphics{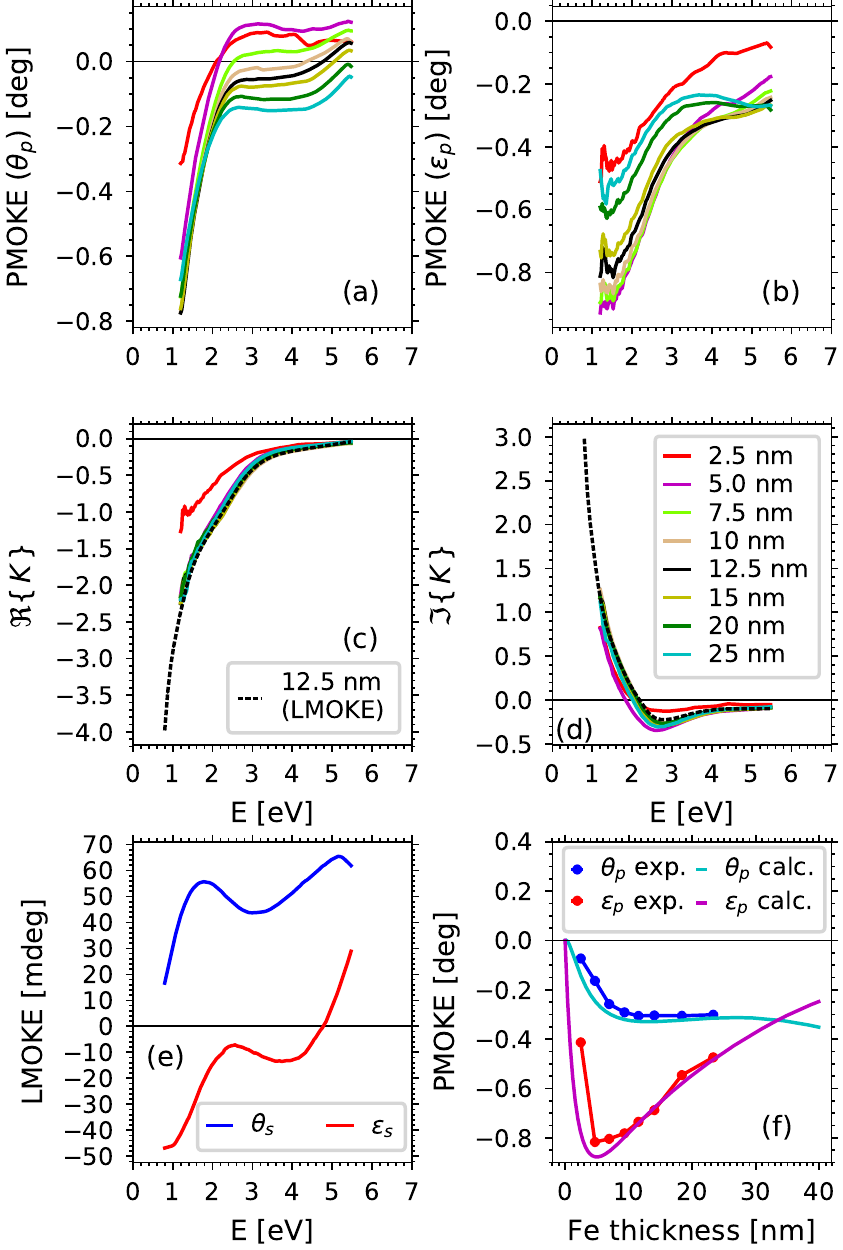}
\end{center}
\caption{Experimental PMOKE spectra of (a) Kerr rotation $\theta_p$ and (b) Kerr ellipticity $\epsilon_p$ at AoI=5$^\circ$, scaled to magnetization saturation. Spectra of the (c) real and (d) imaginary part of the MO parameter $K$ yielded from the saturated PMOKE spectra do not differ significantly with the thickness (except for the sample with a nominal thickness of  2.5\,nm). The $K$ spectra provided by LMOKE spectroscopy of the sample with a nominal thickness of 12.5\,nm agree very well with $K$ spectra obtained from the saturated PMOKE spectra. (e) LMOKE spectra of the sample with a nominal thickness of 12.5\,nm. (f) Thickness dependence of PMOKE at a photon energy of 1.85\,eV at AoI=5$^\circ$.}
\label{LinMOKE_spectra}
\end{figure}

 The LinMOKE spectra provide the spectral dependence of $K$ (after processing by Yeh's 4$\times$4 matrix formalism) which is very important for the further QMOKE spectra processing due to the additional $K^2/\varepsilon_d$ contribution as follows from Eqs.~(\ref{QMOKE_sequence_s}) and (\ref{QMOKE_sequence_44}). Also, it is appropriate to provide the complete spectroscopic description of the samples up to the second order in $\bm{M}$ within this paper. 
 
The PMOKE spectra for all the samples are presented in Figs.~\ref{LinMOKE_spectra} (a, b). In the Figs.~\ref{LinMOKE_spectra}(c, d), we present the $K$ spectra obtained from the PMOKE spectra and in the case of the sample with a nominal thickness of 12.5\,nm from the LMOKE spectra, as well. The  LMOKE spectra are presented in Fig.~\ref{LinMOKE_spectra}(e). It should be noted that the PMOKE spectra were measured with the magnetic field of 1.2~T which is not enough to magnetically saturate the samples out-of-plane. Nevertheless, the PMOKE spectra multiplied by a factor of 2.2 yield spectra in excellent agreement with the $K$ spectra from the LMOKE spectroscopy, both measured on the sample with the nominal thickness of 12.5~nm. We find this excellent agreement as the confirmation of the correctness of the determination of the optical constants of $\varepsilon_d$ and $K$ from experimental data. Note that all the presented spectra in the following Section~\ref{comparison_abini} are recorded only from MOKE measurements with in-plane magnetization, where the samples were always magnetically saturated.
 
Finally, the dependence of the PMOKE scaled to the magnetization saturation on the Fe layer thickness at a photon energy of 1.85\,eV is shown in Fig.~\ref{LinMOKE_spectra}(f). The experimental data follows the predicted dependence well. All the values that were needed for the Yeh's 4$\times$4 matrix formalism were taken from the sample with a nominal thickness of 12.5\,nm and only the thickness of the Fe layer was varied to obtain the thickness dependence (the value of $K$ was provided by LMOKE spectroscopy, hence the experimental value at a nominal thickness of 12.5\,nm does not absolutely follow predicted amplitude as one can notice in Fig.~\ref{LinMOKE_spectra}(f)). A small disagreement between other experimental and calculated values is due to both slightly different $\varepsilon_d$ and $K$ for different Fe thicknesses, as well as a probable small difference in the scaling factor for different Fe layer thicknesses. 

%%%%%%%%%%%%%%%%%%%%%%%%%%%%%%%%%%%%%%%%%%%
\subsection{Quadratic MOKE spectroscopy}
\label{sec:QMOKE_spectroscopy}

\begin{figure}
\begin{center}
\includegraphics{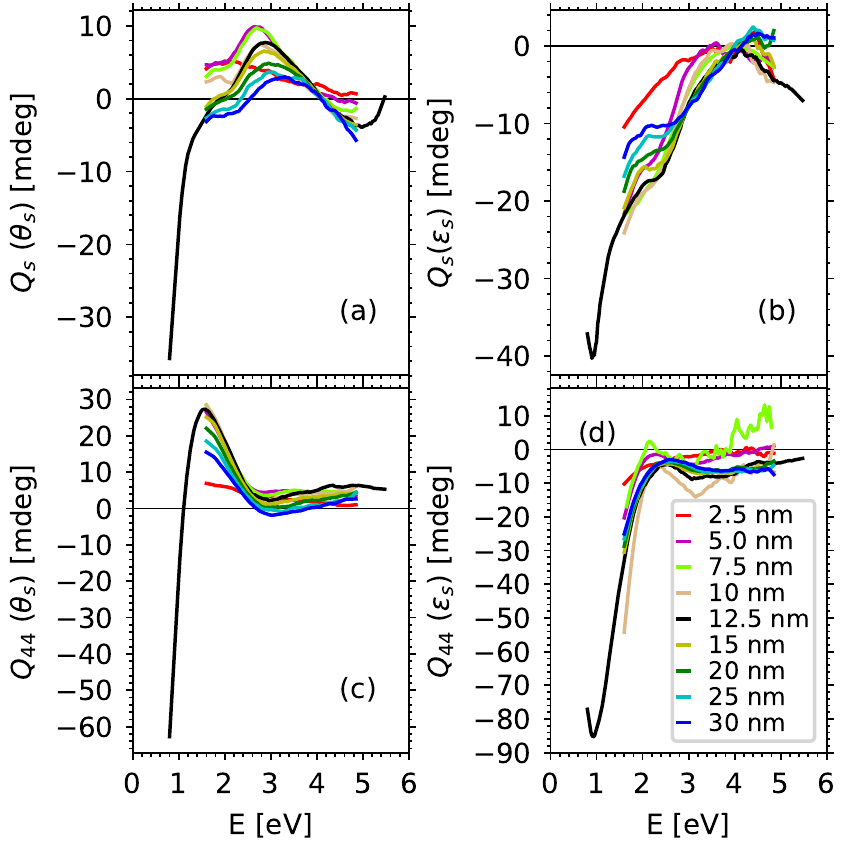}
\end{center}
\caption{(a) Rotation and (b) ellipticity of $Q_s$ spectra. (c) Rotation and (d) ellipticity of $Q_{44}$ spectra. All of them measured with $s-$ polarized incident light. The measured spectra were digitally processed (smoothed) with a Savitzky-Golay filter over the photon energy to improve signal-to-noise ratio. The sample with a nominal thickness of 12.5\,nm was measured with an extended spectral range of 0.8 -- 5.5\,eV. Again, the thinnest sample with a nominal thickness of 2.5\,nm shows the largest deviation compared to the other samples of the thickness dependent series.}
\label{QMOKE_spec}
\end{figure}

The QMOKE spectra for all the samples were measured according to Eqs.~(\ref{QMOKE_sequence_s}) and (\ref{QMOKE_sequence_44}). The measured spectra in the range of 1.6 -- 4.8~eV are presented in Fig.~\ref{QMOKE_spec}. The sample with a nominal thickness of 12.5\,nm was measured at the setup with an extended spectral range of 0.8 -- 5.5~eV. Recall, measured QMOKE also has a contribution from the linear term $K$, being proportional to $K^2/\varepsilon_d$ $M_L M_T$ provided by cross terms $\varepsilon_{yz} \varepsilon_{zx}/\varepsilon_d$ and $\varepsilon_{zy} \varepsilon_{xz}/\varepsilon_d$ (Eq.~(\ref{Kerr_analyt})). Let us emphasize, this quadratic-in-magnetization contribution to MOKE arises from optical interplay of two off-diagonal permittivity elements, both being linear in magnetization.

\begin{figure}
\begin{center}
\includegraphics{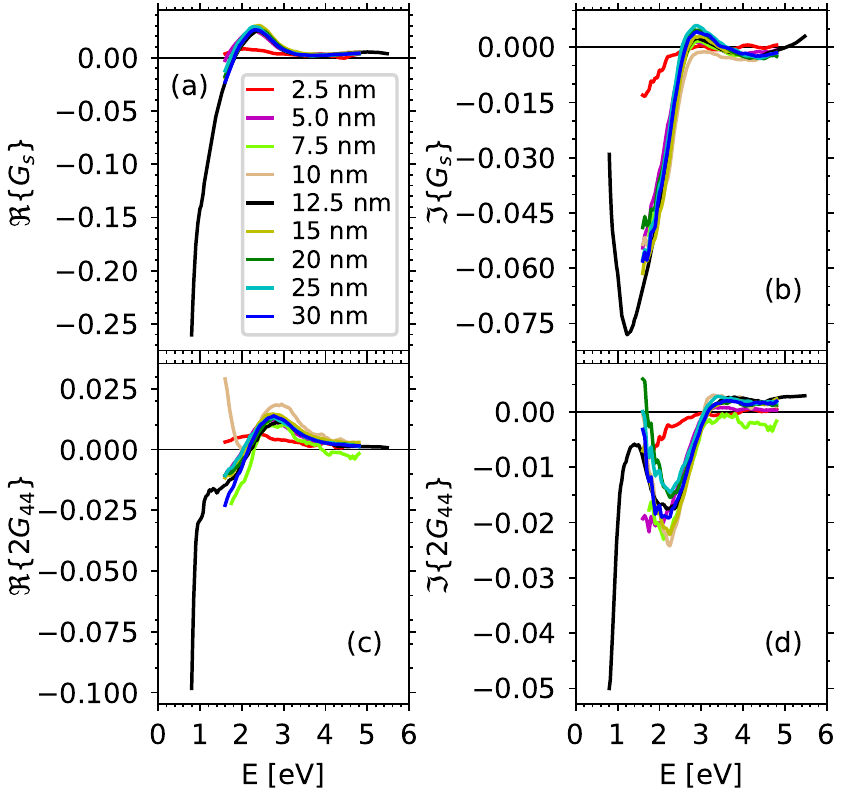}
\end{center}
\caption{Spectra of the (a) real and (b) imaginary part of the quadratic MO parameter $G_s$  and the (c) real and (d) imaginary part of the quadratic MO parameter $2G_{44}$ for all the samples of the series.}
\label{G_MO_par}
\end{figure}

The deduced spectra of the quadratic MO parameters $G_s$ and $2G_{44}$ are shown in Fig.~\ref{G_MO_par}. The shape of the spectra do not substantially change with the thickness, showing that there is no substantial contribution from the interface. 

The only exception (apart from the sample with a nominal thickness of 2.5\,nm, which is also deviating in all previous measurements) is the real part of the $2G_{44}$ spectra below 2\,eV for the sample with a nominal thickness of 10\,nm. The source of this deviation stems from the interplay of two sources: (i) the ellipticity of $Q_{44}$ spectra is almost twice large in the case of this sample, compared to others (see Fig.~\ref{QMOKE_spec}(d)). (ii) The value of $K$ is above 1 in spectral range below 2\,eV (for both the real and imaginary part, and in the absolute value). Thus, the contribution of $K^2/\varepsilon_d$ is the dominant contribution to $Q_{44}$ spectra below 2\,eV, and therefore a small change in the $Q_{44}$ spectra will substantially affect the yielded $2G_{44}$ spectra.  

In Appendix~\ref{MBE_sample} we further present a comparison of $K$, $G_s$ and $2G_{44}$ spectra of the sample with a nominal thickness of 12.5\,nm (prepared by magnetron sputtering) and the sample prepared by MBE. Analogous instability of the $2G_{44}$ parameter can actually be observed here as well. A rather small difference in yielded $K$ spectra and measured $Q_{44}$ spectra provides a significant change of result in yielded $2G_{44}$ spectra. Otherwise, the spectra of samples grown by two different techniques follow the same qualitative progress, but deviate slightly in the magnitude, probably due to small differences in the crystalline quality.

\begin{figure}
\begin{center}
\includegraphics{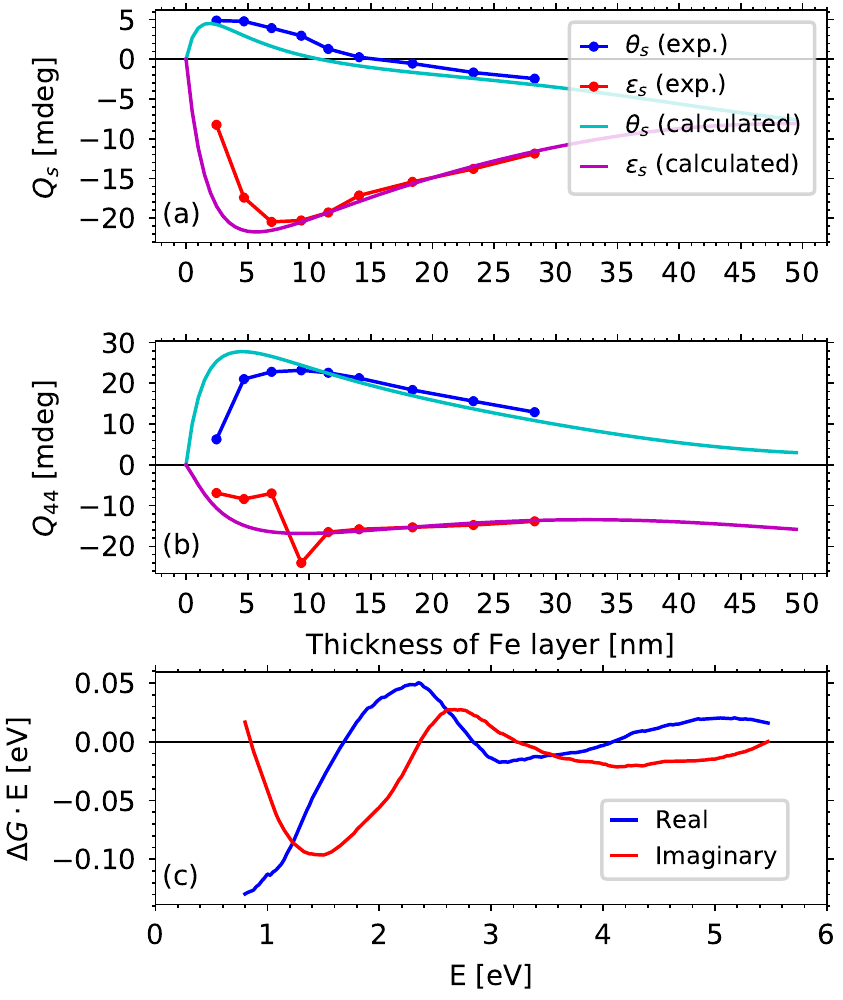}
\end{center}
\caption{Thickness dependence of (a) $Q_s$ and (b) $Q_{44}$ for a photon energy of 1.85\,eV. Lines were provided by Yeh's 4$\times$4 matrix calculus. Both thickness dependencies are for AoI = 5$^\circ$. (c) The spectral dependence of the real and imaginary part of $\Delta G = G_s-2G_{44}$ represents the anisotropy strength of the quadratic MO tensor across the whole spectral range. Every point is weighted by its photon energy for clarity.}
\label{DG_thickness}
\end{figure}

In Figs.~\ref{DG_thickness} (a) and (b) we present the measured and calculated Fe layer thickness dependence for $Q_s$ and $Q_{44}$, respectively. The dependence is for a photon energy of 1.85\,eV and the calculations are provided by Yeh's 4$\times$4 matrix formalism with AoI=5$^\circ$ (being the AoI used within the experiment), where $\varepsilon_d$, $K$, $G_s$ and $2G_{44}$ were taken from the sample with a nominal thickness of 12.5\,nm. The theoretical dependence slightly differs from experimental results for thinner Fe layers. This could be explained by slightly different $\varepsilon_d$, $K$, $G_s$ and $2G_{44}$ for the thinner samples as shown in Figs.~\ref{Elipso}, \ref{LinMOKE_spectra} and \ref{G_MO_par}, respectively, as well as slightly different material properties of capping layers in each sample. Strong deviation could be seen in the case of the experimental value of $Q_{44}$ for the sample with a nominal thickness of 10\,nm, as already discussed above.

The parameter $\Delta G = G_s - 2G_{44}$ provides information about the anisotropy strength of the quadratic MO tensor $\bm{G}$ \cite{Hamrlova2013}. Its spectral dependence for the sample with a nominal thickness of 12.5\,nm is presented  in Fig.~\ref{DG_thickness} (c), shown in the form $\Delta G\cdot E$, i.e.\ multiplied by photon energy.

%%%%%%%%%%%%%%%%%%%%%%%%%%%%%%%%%%%%%%%%%%%
%% ----------COMPARISON
%%%%%%%%%%%%%%%%%%%%%%%%%%%%%%%%%%%%%%%%%%%
\section{Comparison of experimental spectra with calculations and the literature}
\label{comparison_abini}

\begin{figure}
\begin{center}
\includegraphics{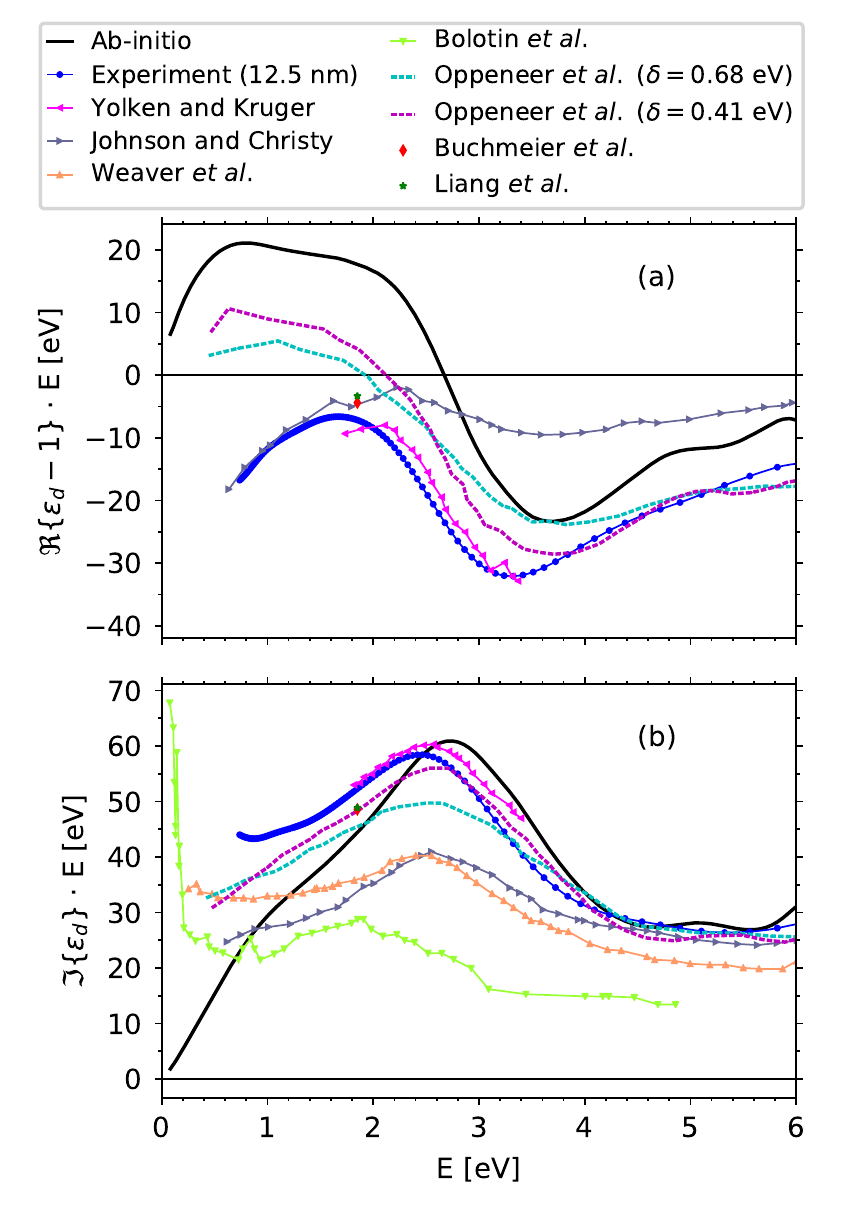}
\end{center}
\caption{Experimental (markers) and ab-initio calculated interband spectra (lines) of (a) real and (b) imaginary part of $(\varepsilon_d-1)\cdot E$ [eV]. Experimental spectra acquired in this work have marker every 10 experimental points (blue bullets). The remaining spectra are taken from literature \protect\cite{Oppeneer92, Yolken65, Johnson73,Buchmeier2009,Liang2015}.}
\label{eps0_comp}
\end{figure}

\begin{figure}
\begin{center}
\includegraphics{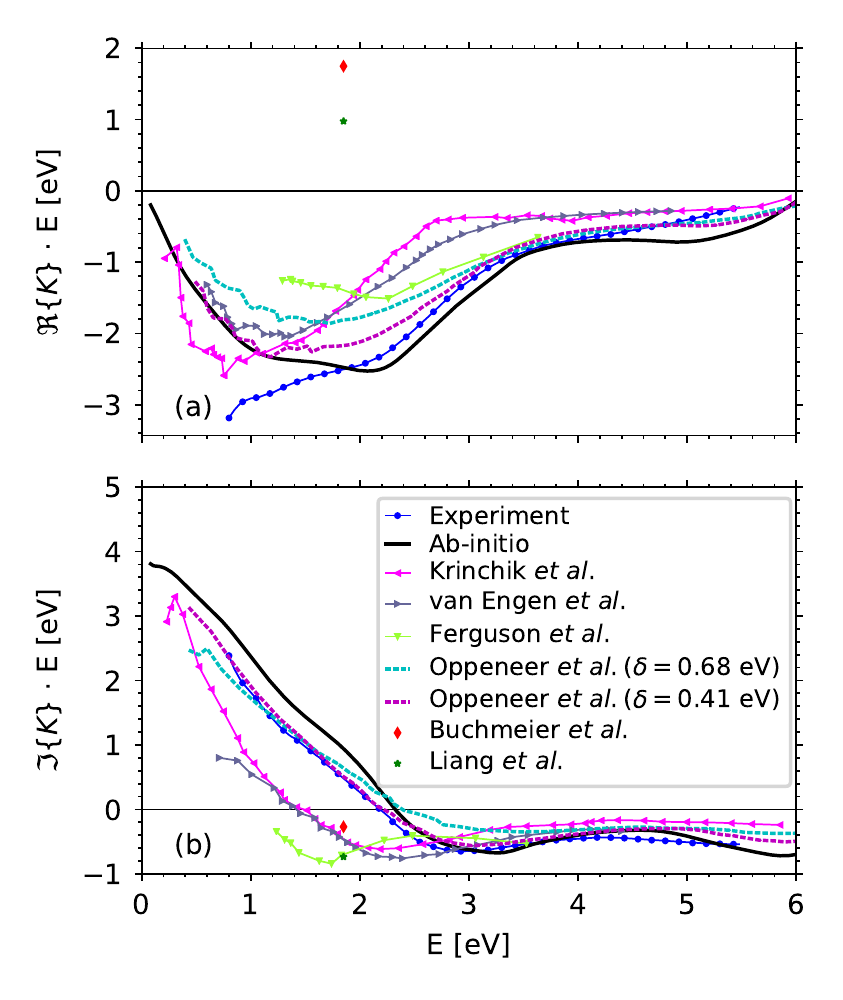}
\end{center}
\caption{Experimental (markers) and ab-initio calculated interband spectra (lines) of the (a) real and (b) imaginary part of $K\cdot$E [eV]. Experimental spectra acquired in this work have a marker every 5 experimental points (blue bullets). The remaining spectra are taken from the literature \protect\cite{Oppeneer92, Krinchik68, Ferguson69,Buchmeier2009,Liang2015}.}
\label{K_comp}
\end{figure}

In this Section, we discuss the comparison of experimental spectra with ab-initio calculations and the literature. All the representative experimental data within this section are from the sample with a nominal Fe thickness of 12.5\,nm. Further, all the spectra in this section are expressed in the form multiplied by photon energy $E$, being an alternative expression of the conductivity spectra. Note that this is analogous to the well-known relation of conversion between complex permittivity and complex conductivity tensor $\varepsilon_{ij}=\delta_{ij}+i\sigma_{ij}\hbar/(\varepsilon_0E)$, where $E$ is the photon energy and $\delta_{ij}$ the Kronecker delta.

The electronic structure calculations of bcc Fe \cite{Stejskal18} were performed using the WIEN2k \cite{Blaha2014} code. The used lattice constant for all calculations was the bulk value, being 2.8665~\AA. The electronic structure was calculated for two $\bm{M}$ directions parallel to Fe[100] and Fe[011], respectively. We used $90^3=729000$ $k$-points in the full Brillouin zone. The product of the smallest atomic sphere and the largest reciprocal space vector was set to $R_{\mathrm{MT}}K_\mathrm{max}=8$ with the maximum value of the partial waves inside the spheres, $l_\mathrm{max}=10$. The largest reciprocal vector in the charge Fourier expansion was set to $G_\mathrm{max}=12$\,Ry$^{1/2}$. The exchange correlation potential LDA was used within all calculations. The convergence criteria were $10^{-6}$\ electrons for charge convergence and $10^{-6}$\,Ry=$1.4 10^{-5}$\,eV for energy convergence.  % -cc 0.000001 -ec 0.000001
The spin-orbit coupling is included in the second variational method.

The Fermi level was determined by temperature broadened eigenvalues using broadening 0.001\,Ry (0.014\,eV). The optical properties were determined within electric dipole approximation using the Kubo formula \cite{Oppeneer92, Draxl2006}. The Drude term (intraband transitions) is omitted in the ab-initio calculated optical and MO properties. We discuss possibilities of how to handle the Drude contribution in Appendix~\ref{s:drude}. By broadening the spectra and applying Kramers-Kronig relations, we obtain a full permittivity tensor $\bm{\varepsilon}$ for each direction of $\bm{M}$. The spectra for $K$, $G_s$ and $2G_{44}$ are obtained directly from the permittivity tensors $\bm{\varepsilon}$ \cite{Hamrlova2016}.

\begin{subequations}
\begin{align}
K&=\frac{1}{2}\left(\varepsilon_{yz}^{([100])}-\varepsilon_{zy}^{([100])}\right),
\label{ab-ini_K}
 \\[1mm]
G_s&=\varepsilon_{xx}^{([100])}-\varepsilon_{yy}^{([100])}, 
\label{ab-ini_Gs}
\\[1mm]
2G_{44}&=\varepsilon_{yz}^{([011])}+\varepsilon_{zy}^{([011])},
\label{ab-ini_G44}
\end{align}
\end{subequations} 
\noindent
where the superscript denotes the $\bm{M}$ direction in the crystallographic structure. 

\begin{figure*}
\includegraphics{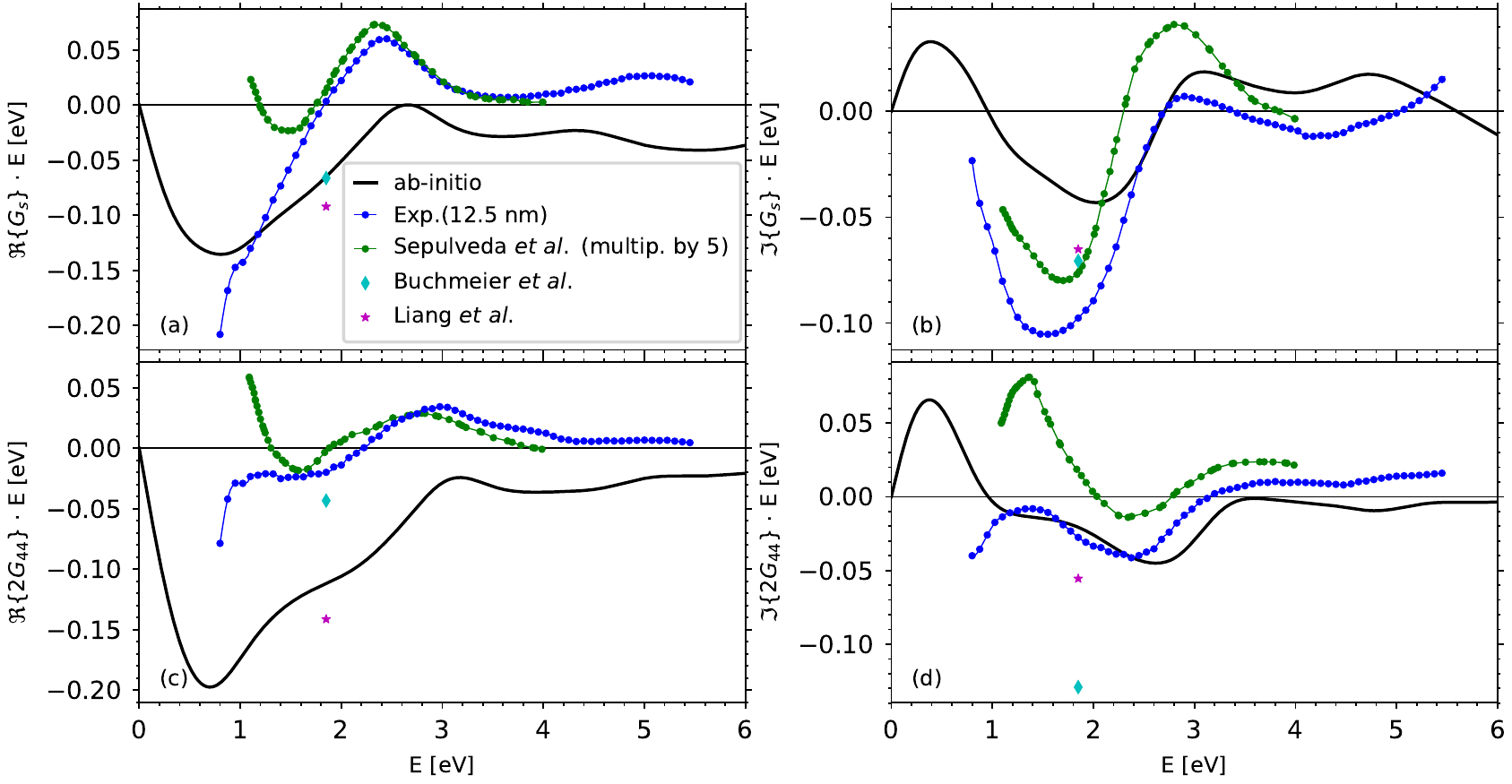}
\caption{The experimental $G_s$ spectra (a, b) and the experimental $2G_{44}$ spectra (c,d) compared with the ab-initio calculations. $G_s$ spectra are calculated for $\vec{M}\parallel[100]$  and $2G_{44}$ spectra calculated for $\vec{M}\parallel[011]$ both with smearing of FWHM=1.2\,eV and with $90^3=729000$ $k$-points in the full Brillouin zone. Further, we show a comparison with data taken  from the literature \protect\cite{Buchmeier2009,Liang2015,Sepulveda2003}. The spectra taken from Sep\'{u}lveda \textit{et al.}\protect\cite{Sepulveda2003} have been multiplied by a factor of 5 to be comparable with our experimental spectra.}
\label{f:GvsAb_ini}
\end{figure*}

Figures~\ref{eps0_comp}(a) and (b) present experimental spectra of $\varepsilon_d-1$  compared to their ab-initio calculations. We also present experimental data from the literature \cite{Yolken65, Johnson73, Oppeneer92} and ab-initio calculations by Oppeneer \textit{et al.} \cite{Oppeneer92} in the same figure. The imaginary (absorption) part of the diagonal permittivity, $\Im(\varepsilon_d)$ is dominated by the absorption peak at 2.4\,eV. This peak originates from transitions of mostly-3d down electrons above and below the Fermi level. The ab-initio calculated peak position is very stable regarding small changes of the lattice constant, magnetization direction, and small distortion of the Fe lattice. On the other hand, the peak position is determined by the selected exchange potential, where LDA provides the closest match to the experimental results, while other potentials (GGA, LDA+U, GGA+U) display larger deviation from the experimental peak position. Therefore we choose the LDA exchange potential to calculate the electronic structure of bcc Fe, although LDA still overestimates the width of the occupied 3d bands. The width of the occupied 3d bands can be corrected using dynamical mean-field theory (DMFT) \cite{Miura2008}. Further, note that the peak amplitude depends on the smearing parameter \cite{Oppeneer92}, and we chose smearing $\delta=0.6$\,eV in the case of $\varepsilon_d$ to adjust the peak height. 

Figures~\ref{K_comp} (a) and (b) show a comparison between experimental and ab-initio calculated spectra of $K$, demonstrating excellent agreement. Note the absorption part corresponds to $\Re(K)$, with two peaks at 2.0 and 1.1\,eV. The amplitude of $\Re(K\cdot E)$ is about -2.5\,eV, i.e.\ about 4\% of the maximal value of $\Im(\varepsilon_d\cdot E)$ being about 60\,eV. Although in both figures (Figs.~\ref{eps0_comp} and \ref{K_comp}) absolute values differ by dozens of percent for some photon energies, the peaks and courses of spectra, being characteristic for the given material, are very similar for all the presented data, both experimental and theoretical (note that disagreement with the reported values at single wavelength \cite{Buchmeier2009, Liang2015} is probably due to sign inconsistency). Further, the d.c. limit of the imaginary part of the $K$ spectra corresponds to the anomalous Hall conductivity. Its value extracted from the ab-initio calculation is 512 ($\Omega$cm)$^{-1}$ (760 ($\Omega$cm)$^{-1}$ without broadening) agreeing with the value provided in Ref.\cite{Yao04}. Finally, note that sign of ab-initio (Wien2k ver. 17) calculated $K$-spectra is reversed, to agree with the sign of the experimental $K$-spectra (this sign error was corrected in Wien2k ver. 19.1).

Figure \ref{f:GvsAb_ini} shows experimental spectra of the real (a) and imaginary (b) part of $G_s$ spectra, compared with the ab-initio calculations. The fundamental (imaginary) part of $G_s$ has a pronounced peak at 1.6\,eV with the amplitude in the experimental spectra being $\Im(G_s\cdot E)=-0.11$\,eV. The main features of $G_s$ are well-described by ab-initio spectra. However, the ab-initio calculated peak at 1.6\,eV has about half that amplitude. Figures \ref{f:GvsAb_ini} (c) and (d) show the real and imaginary part of the experimental spectra of $2G_{44}$, respectively, compared to the ab-initio calculations. In the case of the fundamental part of $2G_{44}$ spectra, both shape and amplitude are well described ab-initio. The larger disagreement between $\Re(G_s)$, $\Re(2G_{44})$  and their ab-initio descriptions (particularly for small photon energies) could be due to the missing Drude term, which is omitted in the ab-initio calculations, and which mainly contributes to the real part of the permittivity at small photon energies. Finally, note that in the ab-initio calculations, convergence (for example on density of the $k$-mesh) of $2G_{44}$ is much better compared to $G_s$, as $G_s$ is calculated as a small change of the diagonal permittivities (Eq.~(\ref{ab-ini_Gs})) whereas $2G_{44}$ is calculated from off-diagonal permittivity (Eq.~(\ref{ab-ini_G44})).

Further, we show the comparison of the spectral dependence of $G_s$ and $2G_{44}$ from Sep\'{u}lveda \textit{et.al.} \cite{Sepulveda2003}. The spectra had to be multiplied by a factor of 5 to be comparable to our experimental and the ab-initio spectra. Then, the agreement is perfect for the real part of both $G_s$ and $2G_{44}$ in the spectral range 1.5--4.0\,eV. The disagreement of spectral dependence under 1.5\,eV can be explained by different sample quality; as the same behaviour was already experienced for $2G_{44}$ in the case of the sample with a nominal thickness of 10\,nm and also in the case of the sample prepared by the MBE, which is discussed in a previous section and in Appendix \ref{MBE_sample}, respectively. The comparison of the imaginary part of $G_s$ and $2G_{44}$ between our data and the scaled data of Sep\'{u}lveda \textit{et al.} \cite{Sepulveda2003} provide very similar behaviour except for some offset and also different amplitude of peaks, especially in case of the $\Im(2G_{44})$ peak at 1.5\,eV. We do not know wherefrom the scaling factor 5 between our data and data of Sep\'{u}lveda \textit{et.al.} is stemming. In the case of Sep\'{u}lveda \textit{et.al.} the data were obtained from experimental measurement of variation of reflectivity with quadratic dependence on magnetization. The poor quality of the samples can be ruled out, as in the case of polycrystalline material $\Delta G = 0$, i.e. $G_s=2G_{44}$, which is not the case here. However, note that our optical spectra of $\varepsilon_d$, $K$, $2G_{44}$ and $G_s$ well describe their experimental reflectivity spectra using our numerical model.

%%%%%%%%%%%%%%%%%%%%%%%%%%%%%%%%%%%%%%%%%%%
%---CONCLUSION
%%%%%%%%%%%%%%%%%%%%%%%%%%%%%%%%%%%%%%%%%%%
\section{Conclusion}

We provided a detailed description of our approach to the QMOKE spectroscopy, which allows us to obtain quadratic MO parameters in the extended visible spectral range. The experimental technique stems from the 8-directional method that separates linear and quadratic MOKE contributions. 

The quadratic magnetooptic parameters $G_s$ and $2G_{44}$ of bcc Fe (expressing magnetic linear dichroism of permittivity along the [100] and [110] directions, respectively) were systematically investigated. The spectral dependence of $G_s$ and $2G_{44}$ is experimentally determined in the spectral range 0.8 -- 5.5\,eV, being acquired by QMOKE spectroscopy and numerical simulations using Yeh's 4$\times$4 matrix formalism. A sample series of Fe thin films with varying thicknesses grown by magnetron sputtering on MgO(001) substrates and capped with 2.5\,nm of silicon were used. Except for the sample with a nominal thickness of 2.5\,nm, the dependence of the obtained spectra on the Fe layer thickness is small, indicating a small contribution of the interface. During our investigations, the linear MO parameter $K$ in the spectral range 0.8 -- 5.5\,eV and the diagonal permittivity $\varepsilon_d$ in the spectral range 0.7 -- 6.4\,eV were also acquired.

Further, all measured permittivity spectra are compared to ab-initio calculations. The shapes of those spectra are well described by electric dipole approximation, with the electronic structure of bcc Fe calculated using DFT with LDA exchange-correlation potential and with spin-orbit coupling included. However, to describe $G_s$ and $2G_{44}$, a fine mesh of 90$\times$90$\times$90 is used as $G_s$ is calculated as a small variation of diagonal permittivity $\varepsilon_{ii}$ with magnetization direction.

With the measurement process well established, the technique is ready to be used on other ferromagnetic materials, and also tested on antiferromagnetic materials. A suitable candidate could be the easy-plane AFM NiO grown on a ferri- or ferromagnetic support in order to control the AFM by the exchange coupling to the additional ferri- or ferromagnetic layer which can be magnetically aligned by an external field. In such a bilayer, the contribution of the ferri- or ferromagnetic layer has to be studied separately in the same manner as we have done here for bcc Fe.

%%%%%%%%%%%%%%%%%%%%%%%%%%%%%%%%%%%%%%%%%%%
%---ACKNOWLEDGEMENT
%%%%%%%%%%%%%%%%%%%%%%%%%%%%%%%%%%%%%%%%%%%
\begin{acknowledgments}
The authors thank G\"{u}nter Reiss,  Jarom\'{i}r Pi\v{s}tora, Gerhard G\"{o}tz and John Cawley for support, assistance and discussion. This work was supported by Czech Science Foundation (19-13310S) and the Deutsche Forschungsgemeinschaft (DFG Re 1052/37-1). The work was also supported by the European Regional Development Fund through the IT4Innovations National Supercomputing Center - path to exascale project, project number CZ. $02.1.01/0.0/0.0/16\_013/0001791$ within the Operational Programme Research, Development and Education and supported in part by OP VVV project MATFUN under Grant CZ.$02.1.01/0.0/0.0/15\_003/0000487$.
\end{acknowledgments}

%%%%%%%%%%%%%%%%%%%%%%%%%%%%%%%%%%%%%%%%%%%
%---APPENDIX
%%%%%%%%%%%%%%%%%%%%%%%%%%%%%%%%%%%%%%%%%%%
\appendix
\section{Sign conventions}
\label{app_conv}
Within the fields of optics and magnetooptics, there is a vast amount of conventions. As there is no generally accepted system of conventions, we define here all conventions adopted within this work.

To describe reflection from a sample, three Cartesian systems are needed, one for incident light beam, one for reflected light beam and one for the sample. All those Cartesian systems are right-handed and defined in Fig.~\ref{xyz_sys} of the main text.
\begin{description}
	\item[Time convention]\quad \newline
	 The electric field vector of an electromagnetic wave is described by negative time convention as $\bm{E}{(\bm{r},t)}=\bm{E}(\bm{r}) e^{-i\omega t}$, providing permittivity in the form $\varepsilon=\Re(\varepsilon) +i\Im(\varepsilon)$, where the imaginary part of complex permittivity $\Im(\varepsilon) >0$.
	\item[Cartesian  referential  of  the  sample] \quad \newline
	The Cartesian system describing the sample is the right-handed $\hat{x}$, $\hat{y}$, $\hat{z}$ system, where $\hat{z}$-axis is normal to the surface of the sample, and points into the sample. The $\hat{y}$-axis is  parallel with the plane of light incidence and with the sample surface, while its positive direction is defined by the direction of $k_y$, being the $\hat{y}$-component of the wave vector of incident light as shown in Fig.~\ref{xyz_sys}. In this system, rotations of the crystallographic structure and magnetization take place.

	\item[Cartesian  referential of light ] \quad \newline
	We use the right-handed Cartesian system $\hat{s}$, $\hat{p}$, $\hat{k}$ for description of the incident and reflected light beam. The direction of vector $\hat{k}$ defines the direction of propagation of light. Vector $\hat{p}$ lies in the incident plane, i.e.\ a~plane defined by incident and reflected beam. The vector $\hat{s}$ is perpendicular to this plane and corresponds to $\hat{x}$.  This convention is the same for both incident and reflected beams (Fig.~\ref{xyz_sys}).
	
	\item [Convention of the Kerr angles]\hfill\newline
	The Kerr rotation $\theta$  is positive if azimuth $\theta$ of the polarization ellipse  rotates clockwise, when looking into the incoming light beam. The Kerr ellipticity $\epsilon$ is positive if temporal evolution of the electric field vector $\bm{E}$ rotates clockwise when looking into the incoming light beam.
	
	\item [\protect{\parbox[b]{7.1cm}{Convention of rotation of the sample, the magnetization and the optical elements}}]\hfill\newline The rotation is defined as positive if the rotated vector pointing in the $\hat{x}$ ($\hat{s}$) direction rotates towards the $\hat{y}$ ($\hat{p}$) direction. The sample orientation $\alpha$ = 0 corresponds to the Fe[100] direction being parallel to the $\hat{x}$-axis and, when looking at the top surface of the sample, the positive rotation of the sample is clockwise. Likewise, the magnetization direction $\mu=0$ corresponds to $\bm{M}$ being in the positive direction of the $\hat{x}$-axis and, when looking at the top surface of the sample, the positive rotation of magnetization $\bm{M}$ is clockwise. Further, when looking into the incoming beam, the positive rotation of the optical elements is counter-clockwise, in contrast to the positive Kerr angles, defined by historical convention.
		
\end{description}

%%%%%%%%%%%%%%%%%%%%%%%%%%%%%%%%%%%%%%%%%%%
%---WRONG SIGN consequences
%%%%%%%%%%%%%%%%%%%%%%%%%%%%%%%%%%%%%%%%%%%
\section{Consequences of the MOKE sign disagreement between the experimental and numerical model}
\label{mag_dir}

The correct sign of LMOKE and QMOKE spectra is given by the conventions used. Nevertheless, to obtain the correct spectra of MO parameters $K$, $G_s$ and $2G_{44}$, the same conventions must be adopted within the numerical model and the experiment. One would intuitively expect only the reversed sign of yielded MO parameters, when the sign conventions of the experiment and the numerical model do not comply. However, completely incorrect values are yielded in this case for the quadratic MO parameters.

 There are numerous points in the experiment where we can go wrong and thus measure the MOKE spectra of the incorrect sign according to our conventions, e.g. wrong direction of in-plane $\bm{M}$ rotation (i.e. $\mu\rightarrow -\mu$), wrong direction of positive external field and thus opposite direction of $\bm{M}$ (i.e. $\mu\rightarrow\mu+180^{\circ}$), error in the calibration process of the setup itself (note that the positive direction of the optical element rotation and the positive direction of the Kerr rotation have opposite conventions) or some quirk in the processing algorithm of the measured data itself (usually we measure change of intensity, which has to be converted to Kerr angles). Further, we can also make a sign error in the code of the numerical model. 

The correct sign of the numerical model output can be checked for by comparison to the simple analytic model that exist for some special cases. E.g.\ PMOKE effect $\Phi$ at the normal angle of incidence for a vacuum/FM(bulk) interface within our sign convention:
\begin{equation}
    \label{eq:PMOKE_analytical}
    \Phi=\frac{\varepsilon_{xy}}{\sqrt{\varepsilon_d^{(FM)}}(1-\varepsilon_d^{(FM)})}
\end{equation}

Various sign mistakes in the experiment will not always lead to the same error, e.g.  the wrong direction of positive external magnetic field will affect the sign of LMOKE spectra but not the QMOKE spectra. On the other hand the wrong direction of $\bm{M}$ rotation will produce a wrong sign of both spectra, LMOKE and QMOKE alike - see the Eqs.(\ref{QMOKE_sequence_s})--(\ref{LMOKE_sequence}). 

In the following, we will discuss a consequence of the latter case, when the direction of $\bm{M}$ rotation has the opposite direction, $\mu'\rightarrow -\mu$ leading to a wrong sign of experimental spectra measured according to Eqs.(\ref{QMOKE_sequence_s})--(\ref{LMOKE_sequence}).  While linear MO parameter $K^{\prime}$, yielded from the LMOKE spectra with a reversed sign, will only have the opposite sign compared to the true MO parameter $K$, the quadratic MO parameters $G_s^{\prime}$ and $2G_{44}^{\prime}$, yielded from the $Q_s$ and $Q_{44}$ spectra with the opposite sign, will be completely different from the true MO parameters $G_s$ and $2G_{44}$, respectively. This is due to the contribution of $K^2/\varepsilon_d$ to the $Q_s$ and $Q_{44}$ spectra, which are invariant to the sign of $K$ itself. Thus, the MO parameters yielded from sign-reversed experimental spectra are bound with the true MO parameters by following equations.

\begin{eqnarray}
K^{\prime}&=&-K\\
G_s^{\prime}&=&-G_s+2\frac{K^2}{\varepsilon_d}\\
2G_{44}^{\prime}&=&-2G_{44}+2\frac{K^2}{\varepsilon_d}	
\end{eqnarray}

\begin{figure}
\begin{center}
\includegraphics{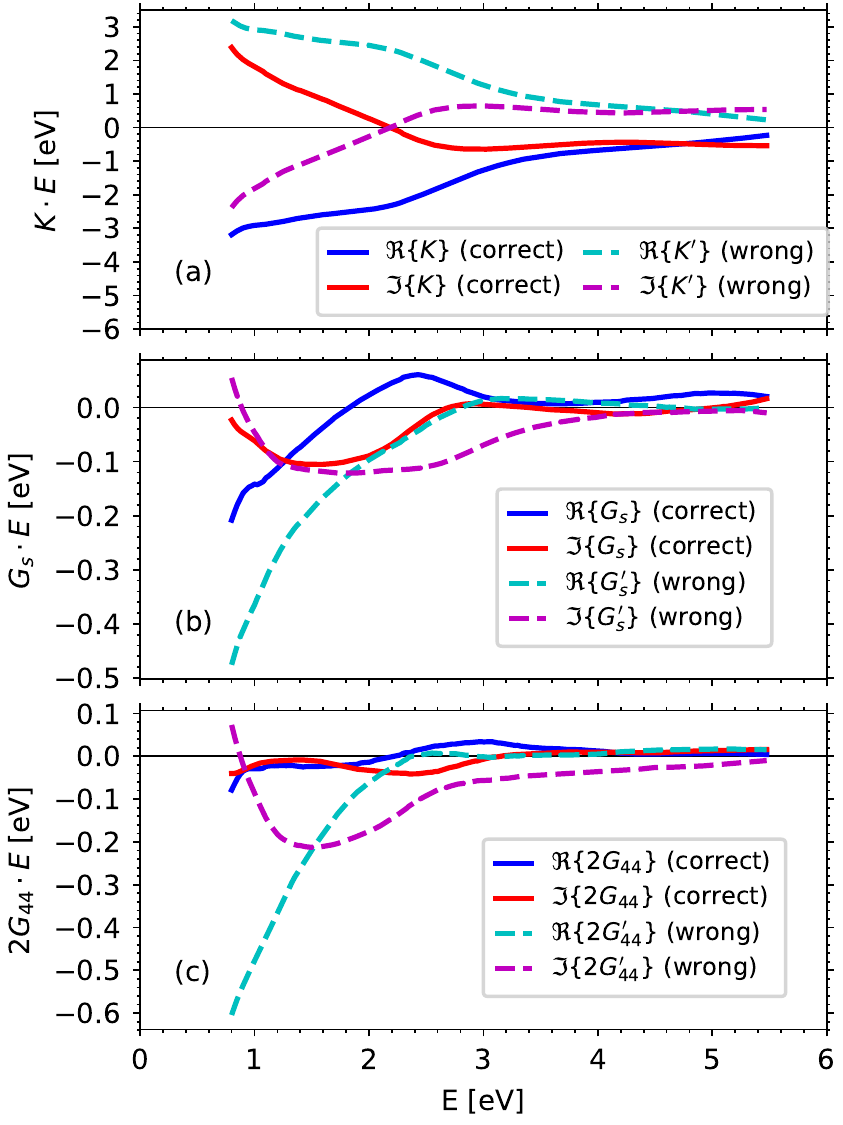}
\end{center}
\caption{Comparison of the spectra of MO parameters of (a) $K$, (b) $G_s$ and (c) $2G_{44}$ to the MO parameters yielded from the experimental spectra with the wrong (reversed) sign of $\mu$ ($K^{\prime}$, $G_{s}^{\prime}$ and $2G_{44}^{\prime}$).}
\label{MOparam_16vs19}
\end{figure}

\noindent
In Fig.~\ref{MOparam_16vs19} we show the wrong MO parameters $K^{\prime}$, $G_s^{\prime}$ and $2G_{44}^{\prime}$  compared to the true MO parameters $K$, $G_s$ and $2G_{44}$.  

Note that neither the shape nor the sign of the true MO parameters is given by the convention used. Any sign conventions can be adopted, but the crucial point is that the conventions used in real experiments and in numerical calculus are the same. Obviously, this issue applies to any error in the experimental setup or the numerical code that would unintentionally reverse the sign of the measured or calculated MOKE spectra, respectively. 

%%%%%%%%%%%%%%%%%%%%%%%%%%%%%%%%%%%%%%%%%%%
%---SPuttering vs. MBE
%%%%%%%%%%%%%%%%%%%%%%%%%%%%%%%%%%%%%%%%%%%
\section{Comparison of the samples grown by molecular beam epitaxy and by magnetron sputtering}
\label{MBE_sample}

Fe and Si films were prepared on a single crystalline MgO(001) substrate via molecular beam epitaxy (MBE). Prior to deposition, the substrates were annealed at 400$^\circ$C for 1h in a 1$\cdot$10$^{-4}$ mbar oxygen atmosphere to remove carbon contamination and obtain defined surfaces. Fe films were deposited by thermal evaporation from a pure metal rod at a substrate temperature of 250$^\circ$C. Silicon capping layers were evaporated at room temperature using a crucible. The deposition rates of 1.89 and 0.3 nm/min for Fe and Si, respectively, were used and controlled by a quartz microbalance next to the source. The base pressure in the UHV chamber was 10$^{-8}$ mbar.

\begin{figure}
\begin{center}
\includegraphics{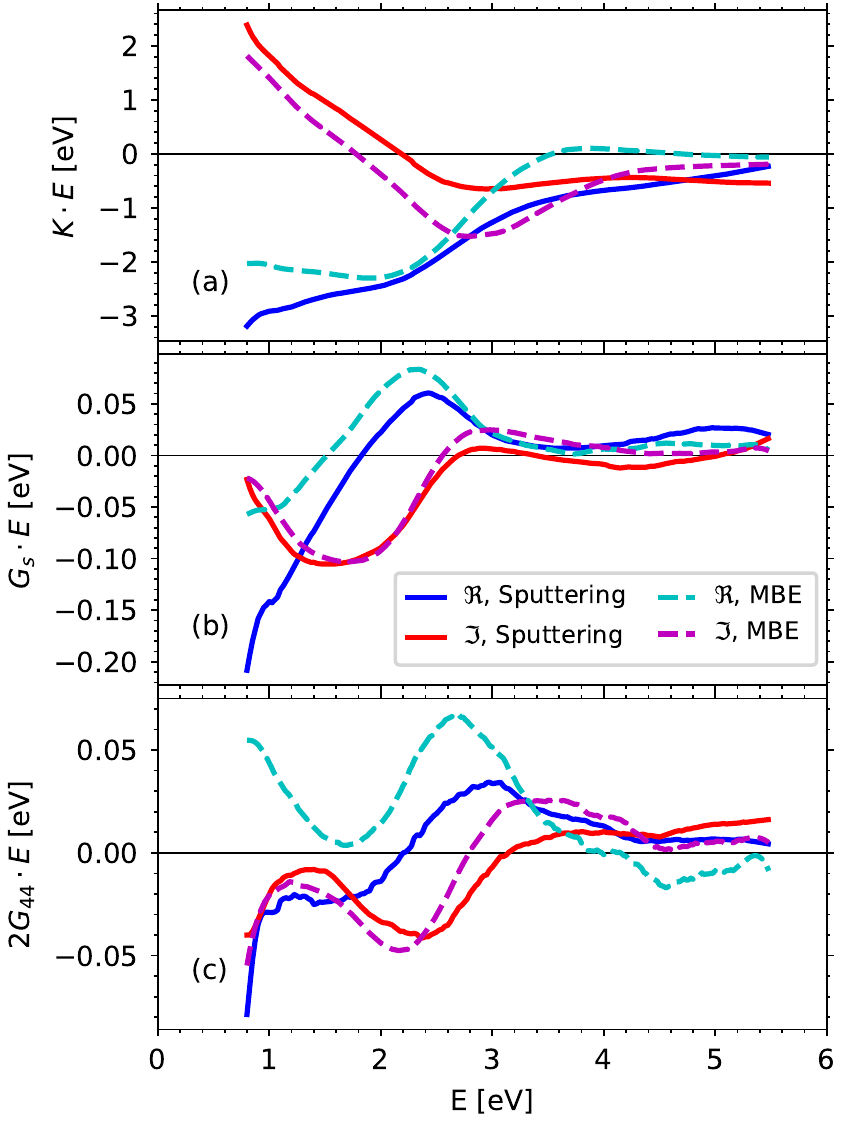}
\end{center}
\caption{Comparison of the spectra of the MO parameters of (a) $K$, (b) $G_s$ and (c) $2G_{44}$ of two samples, one prepared by magnetron sputtering  and the other by MBE. For each of the samples, all the data used within numerical calculations were obtained from the parameters of the particular sample.}
\label{MBE_vs_Sputter}
\end{figure}

The XRD and XRR were measured as described in section~\ref{sample_character}. A thickness of 12.6\,nm was determined by XRR for the MBE prepared Fe layer and 7.0\,nm for the Si+SiO$_x$ capping layer. The thickness of the reference sample with only Si+SiO$_x$ capping was 8.1\,nm. The XRD $\Theta$ -- $2\Theta$ scan was performed around $2\Theta$ = 65$^\circ$ and showed that the samples are of good crystallinity. Further, the ellipsometry, LMOKE and QMOKE spectroscopy were measured on the sample to feed the Yeh's 4$\times$4 matrix calculations with the required sample data. The spectra of $K$, $G_s$, and $2G_{44}$ obtained by numerical calculations are presented and compared to the spectra of the sputter-deposited sample with a nominal thickness of 12.5\,nm in Figs.~\ref{MBE_vs_Sputter}(a)--(c), respectively. The behaviour of the spectra of both samples is very similar, except for the real part of $2G_{44}$ spectra at lower photon energies. Nevertheless the same  discrepancy has already been discussed in the section~\ref{sec:QMOKE_spectroscopy} for the case of the 10\,nm sample. Otherwise the differences of absolute values across spectra are not surprising, as the reported experimental values of MO parameters differ for different samples prepared by different deposition techniques and different groups (as shown in Figs.~\ref{eps0_comp}, \ref{K_comp} and \ref{f:GvsAb_ini}), probably being connected with slightly different crystalline qualities of the Fe layer.

%%%%%%%%%%%%%%%%%%%%%%%%%%%%%%%%%%%%%%%%%%%
%---DRUDE
%%%%%%%%%%%%%%%%%%%%%%%%%%%%%%%%%%%%%%%%%%%
\section{The Drude contribution}
\label{s:drude}

The contribution of intraband transitions in the diagonal permittivity could be described by the classical phenomenological Lorentz-Drude model (in the following, called the Drude term) 
\begin{equation}
\label{eq:drude}
\varepsilon_D=1-\frac{E_p^2}{E^2 + i\Gamma E},
\end{equation}
\noindent
where $E$ is the photon energy, $E_p=\hbar\omega_p$ is the plasma energy describing the strength of the oscillator, with $\omega_p$ being the plasma frequency and $\Gamma=\hbar\tau^{-1}$ the damping constant, and 1 stands for the relative vacuum permittivity.

In order to include the Drude term into $G$-spectra, first recall that $G_s$ and $2G_{44}$ express magnetic linear dichroism (MLD), $G_s=\varepsilon_\parallel-\varepsilon_\perp$ for $\bm{M}\parallel\braket{100}$ and $2G_{44}=\varepsilon_\parallel-\varepsilon_\perp$ for $\bm{M}\parallel\braket{110}$ where parallel ($\parallel$) and perpendicular ($\perp$) denote the direction of the applied electric field (i.e.\ linear light polarization) with respect to the magnetization direction. 
Second, we assume both $\varepsilon_\parallel$ and $\varepsilon_\perp$ are described by the Drude model Eq.~(\ref{eq:drude}), however with plasma energy $E_p$ and damping constant $\Gamma$ slightly different for both $\parallel$ and $\perp$ directions
\begin{equation}
\label{eq:MLD}
\begin{split}
  \mathrm{MLD}_D&=\varepsilon_\parallel-\varepsilon_\perp = 
  \Delta E_p \frac{\partial}{\partial E_p} \varepsilon_D +
  \Delta \Gamma \frac{\partial}{\partial \Gamma} \varepsilon_D 
  \\
  &= \varepsilon_D \left[
    \frac{2\Delta E_p}{E_p} - \frac{i\Delta \Gamma}{-E+i \Gamma}
  \right],
\end{split}
\end{equation}
where $\mathrm{MLD}_D$ denotes the Drude contribution to magnetic linear dichroism, with $\Delta E_p=E_{p,\parallel} - E_{p,\perp}$ and $\Delta \Gamma=\Gamma_\parallel - \Gamma_\perp$ being differences of the plasma energy and the damping constant between parallel and perpendicular magnetization directions, respectively. Due to the anisotropy of $G$-spectra, $\Delta E_p$ and $\Delta\Gamma$ have different values for $\bm{M}\parallel\braket{100}$ and $\bm{M}\parallel\braket{110}$.  

The number of free parameters in the Eqs.~(\ref{eq:drude}) and (\ref{eq:MLD}) can be reduced from four to two if values of d.c.\ conductivity and AMR are known. One part of the sample with a nominal thickness of 12.5\,nm was patterned into a Hall bar with a top down process using UV lithography and Argon milling. Four point conductivity measurements were performed for various applied currents in the $[100]$ direction from 50 to 500\,\text{\textmu A}. The characteristic dimensions of the Hall bar are length = 635\,\text{\textmu m}, width = 80\,\text{\textmu m} and height = 11.5\,nm. The resistivity and thus the conductivity was determined by performing a linear fit to the data. One obtains conductivity values of $\sigma_\parallel=0.4002 \cdot 10^7$~S/m and $\sigma_\perp=0.4020\cdot 10^7$~S/m, being the conductivity with $\bm{M}$ parallel and perpendicular to the current, respectively. The AMR value of 0.45$\%$ correspond well with the literature \cite{Granberg1999}.

%One obtains conductivity values of $\sigma_\parallel=0.4002 \cdot 10^7$~S/m, $\sigma_\perp=0.4020\cdot 10^7$~S/m and $\sigma_0=0.3999\cdot 10^7$~S/m, being the conductivity with $\bm{M}$ parallel, perpendicular to the current and conductivity without specified $\bm{M}$ in the sample, respectively.

The conductivity $\sigma$ and relative permittivity $\varepsilon$ are related by $(\varepsilon-1) \varepsilon_0= i\sigma\hbar / E$, where $\varepsilon_0$ is the vacuum permittivity and $\hbar$ is the reduced Planck constant.
Hence, at $E=0$, d.c.\ conductivity is
\begin{equation}
\label{eq:sigmaGamma}
\sigma=\frac{\varepsilon_0}{\hbar}\frac{E_p^2}{\Gamma}
\end{equation}
and the (d.c.) anisotropy magnetoresistance is 
\begin{equation}
\label{eq:AMR}
  \begin{split}
  \mathrm{AMR} &=\sigma_\perp-\sigma_\parallel
  = \frac{\varepsilon_0}{\hbar} \lim_{E\rightarrow 0} \left[i E(\varepsilon_\parallel-\varepsilon_\perp)\right]
  \\
  & = \sigma \left[ 
  \frac{2\Delta E_p}{E_p} - \frac{\Delta\Gamma}{\Gamma}
  \right].
  \end{split}
\end{equation}

In order to discuss the Drude contribution to $G_s$ and $2G_{44}$, we first determine the Drude contribution to $(\varepsilon_d-1)$. Knowing the experimental value of d.c.\ conductivity and fitting the Drude model of Eqs.~(\ref{eq:sigmaGamma}) and (\ref{eq:drude}) to the difference between experimental and interband (i.e.\ ab-initio) spectra, the only free parameter in the fit is the plasma frequency becoming $E_p=4.95$\,eV and the corresponding damping term $\Gamma=0.082$\,eV. Further, we use the experimental value of d.c.\ AMR. Recall, AMR was measured solely for current in [100], i.e.\ for determination of the Drude contribution to $G_s$. However, we can assume an equal experimental Drude contribution also for the current in the [110] direction when estimating the Drude contribution to $2G_{44}$. Then, combining Eqs.~(\ref{eq:MLD}) and (\ref{eq:AMR}), $\Delta\Gamma$ can be eliminated resulting in only one free parameter $\Delta E_p$ to describe the Drude contribution to $G_{s}$ and $2G_{44}$ spectra. 

\bibliography{qmokefe}

\end{document}